\documentclass[acmsmall]{acmart}
\usepackage{geometry} %
\usepackage{soul}
\usepackage{amsfonts}
\usepackage{amsmath}
\usepackage[framemethod=tikz]{mdframed}
\usepackage{subfiles}
\usepackage{rotating}
\usepackage{subfig}
\usepackage{booktabs}
\usepackage{multirow}

\usepackage{enumitem} % For Roman numerals
\usepackage[linesnumbered,ruled,vlined]{algorithm2e}

\SetCommentSty{mycommfont}
\SetKwRepeat{Do}{do}{while}%
\setcopyright{acmcopyright}
\acmJournal{TDSCI}
\acmYear{2019}\acmVolume{01}\acmNumber{01}\acmArticle{001}\acmMonth{1}
\acmDOI{10.1145/1122445.1122456}
\acmSubmissionID{123-A56-BU3}

\begin{document}
\newcommand{\cSquare}{%\includegraphics[width=0.2cm]{diagrams/square.pdf} 
}
\newcommand{\cTriangle}{%\includegraphics[width=0.2cm]{diagrams/triangle.pdf} 
}
\newcommand{\cCircle}{%\includegraphics[width=0.2cm]{diagrams/circle.pdf} 
}
\newcommand{\cHexagon}{%\includegraphics[width=0.2cm]{diagrams/hex.pdf} 
}
%
% The "title" command has an optional parameter, allowing the author to define a "short title" to be used in page headers.
\title{\textit{k}D-STR: A Method for Spatio-Temporal Data Reduction and Modelling}

%
% The "author" command and its associated commands are used to define the authors and their affiliations.
% Of note is the shared affiliation of the first two authors, and the "authornote" and "authornotemark" commands
% used to denote shared contribution to the research.
\author{Liam Steadman}
% \authornote{Both authors contributed equally to this research.}
\email{L.Steadman@warwick.ac.uk}
% \orcid{1234-5678-9012}
\author{Nathan Griffiths}
\email{Nathan.Griffiths@warwick.ac.uk}
\author{Stephen Jarvis}
\email{S.A.Jarvis@warwick.ac.uk}
\affiliation{%
  \institution{The University of Warwick}
%   \streetaddress{Department of Computer Science, University of Warwick}
  \city{Coventry}
  \state{Warwickshire}
  \postcode{CV4 7AL}
}

\author{Mark Bell}
\email{MBell@trl.co.uk}
\author{Shaun Helman}
\email{SHelman@trl.co.uk}
\author{Caroline Wallbank}
\email{CWallbank@trl.co.uk}
\affiliation{%
  \institution{TRL}
  \streetaddress{Crowthorne House, Nine Mile Ride}
  \city{Wokingham}
  \state{Berkshire}
  \postcode{RG40 3GA}
}

%
% By default, the full list of authors will be used in the page headers. Often, this list is too long, and will overlap
% other information printed in the page headers. This command allows the author to define a more concise list
% of authors' names for this purpose.
\renewcommand{\shortauthors}{Steadman and Griffiths, et al.}

%
% The abstract is a short summary of the work to be presented in the article.
\begin{abstract}
    Analysing and learning from spatio-temporal datasets is an important process in many domains, including transportation, healthcare and meteorology. In particular, data collected by sensors in the environment allows us to understand and model the processes acting within the environment. Recently, the volume of spatio-temporal data collected has increased significantly, presenting several challenges for data scientists. Methods are therefore needed to reduce the quantity of data that needs to be processed in order to analyse and learn from spatio-temporal datasets. In this paper, we present the $k$-Dimensional Spatio-Temporal Reduction method ($k$D-STR) for reducing the quantity of data used to store a dataset whilst enabling multiple types of analysis on the reduced dataset. $k$D-STR uses hierarchical partitioning to find spatio-temporal regions of similar instances and models the instances within each region to summarise the dataset. We demonstrate the generality of $k$D-STR with 3 datasets exhibiting different spatio-temporal characteristics and present results for a range of data modelling techniques. Finally, we compare $k$D-STR with other techniques for reducing the volume of spatio-temporal data. Our results demonstrate that $k$D-STR is effective in reducing spatio-temporal data and generalises to datasets that exhibit different properties.
\end{abstract}

%
% The code below is generated by the tool at http://dl.acm.org/ccs.cfm.
% Please copy and paste the code instead of the example below.
%


%
% Keywords. The author(s) should pick words that accurately describe the work being
% presented. Separate the keywords with commas.

%
% A "teaser" image appears between the author and affiliation information and the body 
% of the document, and typically spans the page. 
%%\begin{teaserfigure}
%%  \includegraphics[width=\textwidth]{sampleteaser}
%%  \caption{Seattle Mariners at Spring Training, 2010.}
%%  \Description{Enjoying the baseball game from the third-base seats. Ichiro Suzuki preparing to bat.}
%%  \label{fig:teaser}
%%\end{teaserfigure}

%
% This command processes the author and affiliation and title information and builds
% the first part of the formatted document.
\maketitle

\section{Introduction}
Spatio-temporal data generated by sensors in an environment is widely used in many domains, and analysing or learning from such data allows us to understand processes in the environment. In spatio-temporal datasets, correlations between nearby sensors and time intervals can be exploited to better model or predict trends \cite{10.2307/143141}. For example, in the transportation domain, Lv \textit{et al}. used spatial and temporal correlations to create a traffic flow prediction model using a deep learning architecture \cite{6894591}. 
Common tasks in analysing or learning from spatio-temporal data include: (i) imputing missing instances at locations or times not sampled; (ii) identifying unusual behaviours, such as sensors that perform unexpectedly or time periods wherein instances do not fit expected trends; (iii) calculating summary statistics over features or calculating the variance from the expected trend within a time period; (iv) comparing time periods or sensors, for example performing a month-on-month time series analysis; and (v) predicting future instances. These tasks are all aided by the correlations present in spatio-temporal data.

In recent years, the volume of spatio-temporal datasets has increased significantly, presenting several challenges for data scientists. 
Such increases in data volume require more computational time and memory to process and analyse the data. 
Often this task can become infeasible, and so methods are required to reduce the volume of data to be processed whilst minimising the error introduced in later analysis or modelling. 
The aim is not necessarily to compress the data, since many analysis tasks cannot be achieved on compressed data. 
Rather, the aim is to summarise the data in a way that allows analysis to be performed directly on the summarised data.

In previous work \cite{2d-str}, we introduced the 2-Dimensional Spatio-Temporal Reduction (2D-STR) method for reducing the quantity of data used to store a spatio-temporal dataset with a single temporal dimension and single spatial dimension, such as a link-based representation for transport infrastructure.
2D-STR partitions the dataset into a set of \textit{spatio-temporal regions}, and models the instances within each region to summarise the data. 
Beginning with a single region, 2D-STR trades improved reconstruction accuracy and increased storage overhead to meet the user's desired level of reduction. 
This is achieved by either increasing the number of partitions, or increasing the accuracy of the model in one of the regions of the reduced dataset. 

However, 2D-STR cannot be applied directly to datasets containing two or more spatial dimensions since it is ambiguous how sensors and time intervals should be grouped into partitions.
Therefore, this paper makes three main contributions:
(i) we introduce $k$D-STR, which provides a solution for this issue and generalises the reduction process for $k$ spatio-temporal dimensions, 
(ii) we demonstrate the generality of $k$D-STR for datasets exhibiting different spatio-temporal characteristics, and
(iii) we provide an analysis of the time and memory complexity of $k$D-STR.

Compared with other model-based reduction techniques, $k$D-STR uses the variability of instances in time and space to partition the dataset, rather than adapting a fixed-size partitioning scheme.
By modelling all instances within each partition, $k$D-STR can capture the nuances of the data whereas existing techniques often replace similar partitions with links to the same model. 
Furthermore, instances can be imputed using just the desired location and time as input for the stored models. 
In contrast, existing techniques require further transformations to the model output, increasing the computation time of information retrieval.

% In this paper, we introduce $k$D-STR, which generalises the approach taken by 2D-STR for $k$ spatio-temporal dimensions. 
% $k$D-STR first discretises space and time into polygons around instances, and then groups these polygons into regions of similar instances. 
% Each region is modelled using an appropriate technique given the desired uses of the reduced dataset. 
% In contrast to 2D-STR, we bound the parameter, $\alpha$, which prioritises minimising storage cost and information loss to (0,1), enabling users to choose a value that reflects their desire and trade-off between storage used and information lost more intuitively. 
% We demonstrate the generality of $k$D-STR with 3 datasets, present results for a range of data reduction scenarios, and compare $k$D-STR with other techniques for reducing a dataset.

The remainder of this paper is structured as follows. 
Section 2 reviews existing methods for reducing the volume of spatio-temporal datasets. 
Section 3 formalises the problem of reducing a dataset in a manner that enables analysis to be performed on the reduced data, and introduces the notation used in the paper. 
Section 4 describes the proposed approach, $k$D-STR.
Sections 5 and 6 evaluate the effectiveness of $k$D-STR on three sources of data exhibiting different spatio-temporal characteristics, and compare its performance with existing reduction techniques.
Finally, Section 7 concludes the paper and gives future directions of this work.

\section{Related Work}
The quantity of data present in many spatio-temporal datasets makes them difficult or infeasible to process in their raw forms. 
To facilitate faster modelling and analysis, several techniques exist for reducing the quantity of data that needs to be processed. 
Such techniques aim to minimise the difference between analysis performed on, or models created from, the original dataset and the reduced dataset. 
Other techniques result in smaller reduced datasets, yet the reduced data output is only useful for answering specific questions.
In this section, we summarise the existing methods for reducing the quantity of data to be processed in a spatio-temporal dataset.

\subsection{Feature Selection and Extraction Techniques}
Many existing methods for reducing datasets focus on removing a subset of features, thereby reducing the quantity of data stored for each instance. 
Feature selection methods, which choose a subset of features from the original dataset, can be separated into three categories. 
First, filter methods rank features according to a relevance criterion, such as Shannon entropy, and remove any features that score below a defined threshold. 
However, choosing such a threshold is often dataset specific and can be time consuming. 
Second, wrapper methods use search algorithms to find the optimal subset of features according to an objective function \cite{CHANDRASHEKAR201416}. 
Third, embedded methods aim to incorporate feature selection as part of the training process of machine learning. 
Embedded methods evaluate different fixed-size subsets of features to find the subset that consistently yields the smallest classification error.

Several feature selection techniques for real-valued data, such as that generated by sensors in urban datasets, have been evaluated and compared in the context of different domains.
Meskina compared the results of building machine learning models on full datasets versus the same datasets after they had been reduced by the FOCUS \cite{Almuallim:1991:LMI:1865756.1865761} and RELIEF \cite{kira1992feature} filtering methods \cite{meskina2013effect}. 
FOCUS iteratively searches increasingly larger combinations of features until a combination is found that accurately separates two target classes, whilst RELIEF ranks features using a statistical relevance measure and selects the fewest ranked features that are sufficient to separate classes. 
Meskina found the outputs of both methods achieved similar accuracy rates for Support Vector Machine, Na\"{i}ve Bayes and k-nearest neighbour classification compared to using the full dataset. 
However, the time required for classification was remarkably lower \cite{meskina2013effect}. 
In a similar evaluation, Christopher and Balamurugan found correlation based feature selection achieves 97.8\% of the accuracy achieved with the entire dataset using 12 of the original 181 features \cite{christopher2013feature}. 
Several other feature selection techniques for real-valued data exist, and a number of reviews of these can be found in the literature \cite{liu2010feature, vergara2014review, 6853478, 7339682, Li:2017:FSD:3161158.3136625, CHANDRASHEKAR201416}.

In contrast to feature selection, feature extraction (or feature engineering) methods project the original features onto a new feature space, often of a differing dimensionality. 
The best mapping is that which optimises an objective criterion, such as explained variance or accuracy, when combined with modelling. 
Linear feature extraction algorithms include Principal Components Analysis (PCA) \cite{pearson1901liii} and Linear Discriminant Analysis (LDA) \cite{908974}. 
Whilst PCA maximises the inter-feature variance for created features, LDA minimises the variance within a class and maximises the variance between classes. 
Assumptions made by PCA often require adaptations for spatio-temporal data, and these have been discussed in literature \cite{demvsar2013principal}.
Non-linear algorithms, sometimes referred to as manifold learning, map high-dimensional datasets to lower dimensions such that the mapping reflects the structural features of the original dataset. 
Examples such as Isomap, which uses a geodesic distance to measure the distance between instances, and Locally Linear Embedding (LLE) \cite{roweis2000nonlinear}, which improves on Isomap by reducing the computation required, have been reviewed and compared in literature \cite{Liu:2017:RCD:3135954.3135965}.

Whilst effective at reducing the volume of a dataset, feature selection and extraction techniques do not take advantage of the correlations and patterns present in spatio-temporal data. 
Furthermore, they may remove features that are significant for subsequent analysis or fail to capture nuances that may be of interest in later processing \cite{pmlr-v4-janecek08a}. 
Removing features requires the user to have knowledge of the analysis they are to perform ahead of time. 
Instead, it may be more beneficial to retain information about all features. 
Therefore, whilst feature selection and extraction techniques are widely used, methods that take advantage of the correlations present in spatio-temporal data and retain information about all features may be more useful.

\subsection{Instance Selection and Abstraction Techniques}
Instance selection techniques choose a subset of instances from the original dataset that are sufficient for later processing tasks. 
For example, in a classification task they choose and retain only those instances that are sufficient to accurately classify unseen data. 
Several important and established algorithms currently exist for this task, such as the IB3 incremental algorithm \cite{Aha1991}. 
In IB3, a set of output (i.e., selected) instances $\mathcal{S}$ is initialised to contain a single instance chosen from the set of input instances. 
Then, the remaining instances in the input set are considered in turn. 
If an instance being considered, $x$, and its nearest neighbour in $\mathcal{S}$ have different class labels, $x$ is added to $\mathcal{S}$. 
However, if the nearest neighbour to $x$ in $\mathcal{S}$ has the same class label, $x$ is disregarded and not added to $\mathcal{S}$.
After all input instances are considered, $\mathcal{S}$ contains the instances that are sufficient for classifying instances similar to the input instances.
In the spatio-temporal domain, Whelan \textit{et al.} have used the k-medoids clustering technique to reduce a dataset to $k$ instances \cite{5541993, 5969100}. 

Like feature extraction, instance abstraction techniques create a smaller set of new \textit{prototype} instances which represent the original instances but do not necessarily exist in the original dataset. 
Prototyping has been shown to create reduced training sets for tasks such as k-nearest neighbour classification, and training models on these reduced datasets is demonstrably faster with minimal effects on classification accuracy \cite{Ougiaroglou:2012:EDS:2371316.2371349}. 
A simple example, the Prototypes for Nearest Neighbour (PNN) algorithm, is a supervised method which iteratively creates weighted prototypes that are expected to achieve approximately the same classification accuracies as the original instances \cite{1672420}. 
Another example, the Decision Surface Mapping (DSM) algorithm, selects instances randomly from the set of original instances to become prototypes \cite{80344}. 
Each of the instances in the original dataset are then considered and classified according to the prototypes. 
When an instance is incorrectly classified, the DSM algorithm rewards the nearest neighbour of the correct class by moving it closer to the considered instance whilst the nearest overall neighbour is moved further away. 
The Learning Vector Quantization (LVQ) family of algorithms operate in a similar fashion to the k-means algorithm \cite{Kohonen:2001:SM:558021}. 
Instead of updating prototypes only when a misclassification is made, the LVQ algorithm updates prototypes even when an instance is correctly classified.
Comparisons of instance abstraction techniques can be found in the literature \cite{5709998}.

Like feature selection methods, both instance selection and abstraction techniques remove instances from the dataset that may be of significance in later processing. 
For example, removing instances may increase inaccuracies in imputation tasks or in the calculation of statistics. 
Furthermore, querying the removed instances may no longer be possible. 
Therefore, techniques such as these restrict the analysis that can be performed on the reduced data they output.

\subsection{Data Sketching}
The techniques discussed until now focus on removing or prototyping features and instances. 
In contrast, data sketching techniques create query-specific summaries of the data using a fixed number of passes over the data. 
% Many of these techniques require just one pass over the data and so are very efficient.
Many sketching techniques focus on counting items, such as the Count-Min sketch and its adaptation for real-valued data \cite{cormode2005improved, 8405715}. 
Another, the Bloom Filter, answers membership questions using hash tables \cite{Bloom:1970:STH:362686.362692}. 
The HyperLogLog (HLL) algorithm uses a probabilistic counter to answer cardinality questions and is sufficiently efficient to be used with very large quantities of data \cite{flajolet2007hyperloglog}. 
However, these techniques do not consider the spatial and temporal nature of the data and do not support analysis questions such as those presented in Section 1.

In the spatio-temporal domain, methods have been proposed that combine instance selection data sketching with the Kalman filter to track large-scale spatio-temporal processes \cite{7893735}. 
Furthermore, Tai \textit{et al.} presented a sketching method for building linear classifiers over a spatio-temporal dataset \cite{Tai:2018:SLC:3183713.3196930}. 
By building linear classifiers over the temporal stream of each sensor, correlated features can be identified whilst also permitting analysis of a stream's instances. 
However, this methodology destroys features which are not heavily weighted by the linear classifier. 

Sketching techniques are limited in the analyses or later modelling they permit \cite{cormode2012synopses}. 
They are created specifically for particular queries and, since the original dataset is destroyed after the sketch is created, it is not possible to reconstruct the data for other analyses. 
Most do not take advantage of the correlations present in spatio-temporal data, and many require knowledge of which features will be of interest before the sketch is created. This may not be known ahead of time.

\subsection{Data Reduction using Modelling}

Whilst the techniques discussed above result in the loss of instances or features, some techniques exist for reducing a dataset using statistical modelling. 
The IDEALEM algorithm partitions a data stream into blocks of a fixed size \cite{Lee:2016:NDR:2949689.2949708, 8085035}. 
Key statistical properties about these blocks, such as min, max and average values, are then used to identify those blocks which are statistically similar. 
For each set of similar blocks, the raw data of one block is retained along with statistics about the block and where it repeats in the data stream. 
By processing each of its prototype blocks, IDEALEM allows us to identify unusual temporal periods that do not fit expected trends.
It also enables comparison of different sensors or time periods, retains information about all features, and allows for the faster generation of statistics compared to the original data stream.
However, replacing blocks with links to a similar block introduces error.
Furthermore, since the method does not consider the spatial nature of spatio-temporal data, IDEALEM does not permit spatial imputation without further modelling.

Similar to IDEALEM, the ISABELA algorithm partitions each feature into fixed size spatial windows and sorts the observed values into ascending order within each window \cite{6114457}. 
A B-spline curve is then fitted to each window and the parameters of the curve stored using temporal encoding. 
Since the instances in each window are stored in ascending order by value, a mapping back to the temporal ordering also needs to be stored. 
This is not necessary with many real-world sensor datasets, such as traffic data, which are more smooth and cyclic than the data motivating ISABELA\footnote{In other work, discrete cosine transforms have been shown to effectively reduce similar datasets \cite{6410152}.}.
Furthermore, the reduced representation given by ISABELA makes imputation of values impossible without mapping the data back into temporal order. 
Statistics can be generated over the data if the temporal period of interest is exactly covered by a window, otherwise mapping back into temporal order is again required.
In the same way, identifying unusual spatial or temporal regions is partially supported.

Deep autoencoders have also been used to model the temporal features of spatio-temporal datasets \cite{7727605}. 
The Sparse Autoencoder (SAE) has been used to reliably estimate missing data in spatio-temporal sensor datasets \cite{Wong:2014:IMV:2641798.2641816}. 
This fitting of a summary, which minimises the root mean square error (RMSE) over instances in the discrete spatial and temporal dimensions, is able to impute missing values given other instances from the same time interval. 
It may be possible to adapt this approach to incorporate multiple time intervals, e.g., the whole dataset, and store the autoencoder weights for the purposes of reproducing the dataset. 
However, autoencoder weights are incomprehensible in analysis and so prevent manual analysis of the reduced dataset.

In the domain of traffic dataset analysis, Pan \textit{et al}. have proposed a two-part algorithm that summarises a spatio-temporal traffic sensor dataset \cite{Pan:2010:SST:1878500.1878504}. 
Their method creates a spatio-temporal signature of the dataset using a technique such as wavelet decomposition, and a set of outliers that fall outside an acceptable error margin of this signature. 
Whilst this technique is good at capturing the cyclic and seasonal natures of some datasets, it performs poorly on datasets containing irregular patterns or many outliers.
For example, in road traffic data, instances from national holidays (temporal domain) and areas of construction work (spatial domain) are known to deviate from regular traffic cycles, and so will be labelled as outliers. 
The algorithm will only retain some of these outlier instances due to its probabilistic nature. 
Therefore, reconstruction of these spatial and temporal regions may be poor.

\section{Spatio-Temporal Data Reduction}
\label{sec:preliminaries}
To capture information that may be needed for later analysis, many datasets sample spatial and temporal processes at high frequency and resolution. This often results in datasets that contain regions of low variability where the process being sampled exhibited little change. For example, many weather datasets are sampled at 15 minute intervals, yet the weather they report can exhibit low change for several hours at a time. Therefore, such spatio-temporal datasets are very large but contain high spatial and temporal autocorrelation.

To decrease the quantity of data to be analysed or modelled in these cases, we can group similar adjacent instances together to form spatio-temporal regions. Each of these regions can then be reduced to a model of the instances within the region. 
This reduces both the number of instances in the reduced dataset, since each region becomes a single model in the reduced dataset, and the quantity of data used to store the information within each region. 
Whilst IDEALEM \cite{Lee:2016:NDR:2949689.2949708, 8085035}, ISABELA \cite{6114457}, and the method presented by Pan \textit{et al}. 
\cite{Pan:2010:SST:1878500.1878504} take a similar approach, they use either single or fixed-sized regions defined over the spatial and temporal domains. 
Instead, by forming regions where the dataset exhibits little change, we argue that less complex models can be used to accurately represent the data.
The nuances of the original instances can be maintained, and answering queries on the reduced dataset is still supported. 
In this section, we formalise this approach and the notation used in this paper.

A spatio-temporal dataset $D$ is a set of instances generated by a set of synchronous or asynchronous sensors. 
Each instance in $D$ is recorded at a location $s \in S^{\mathcal{D}}$ at time $t \in T$. 
Therefore, we can reference each instance in $D$ using the notation $d_{t,s}$. 
Here, $T$ is the continuous 1-dimensional temporal domain and $S^{\mathcal{D}}$ is the continuous $\mathcal{D}$-dimensional spatial domain, thus the spatio-temporal space is $k$ dimensional, where $k = 1 + \mathcal{D}$.
In this work, we assume that each sensor exists at a unique location and records at most one instance at any given time, thus only one instance can exist at a given time and location $(t,s)$\footnote{In some scenarios a dataset may contain multiple instances recorded at the same location and time. For example, multiple sensors may be located at the same location and record instances at the same time. In these scenarios, multiple models may be output by the reduction process, thus allowing multiple response values to be modelled for the same location and time input. This is discussed further in Section \ref{sec:conclusion}.}.

Each instance $d_{t,s}$ is a vector of values over the set of features $F$, i.e., $d_{t,s} = \langle d_{t,s}^1,...,d_{t,s}^{|F|}\rangle$. 
For example, in a weather dataset these features may be rainfall, temperature and humidity. 
For generality we assume that these features are real-valued. 
Therefore, the dataset is a mapping from the $k$-dimensional spatio-temporal space to the $|F|$-dimensional feature space, $D:T \times S^{\mathcal{D}} \rightarrow \mathbb{R}^{|F|}$. 
Techniques exist for representing binary and categorical features as real-valued data, and appropriate partitioning algorithms can be used for binary and categorical features.

\begin{figure}[t]
\includegraphics[width=\textwidth]{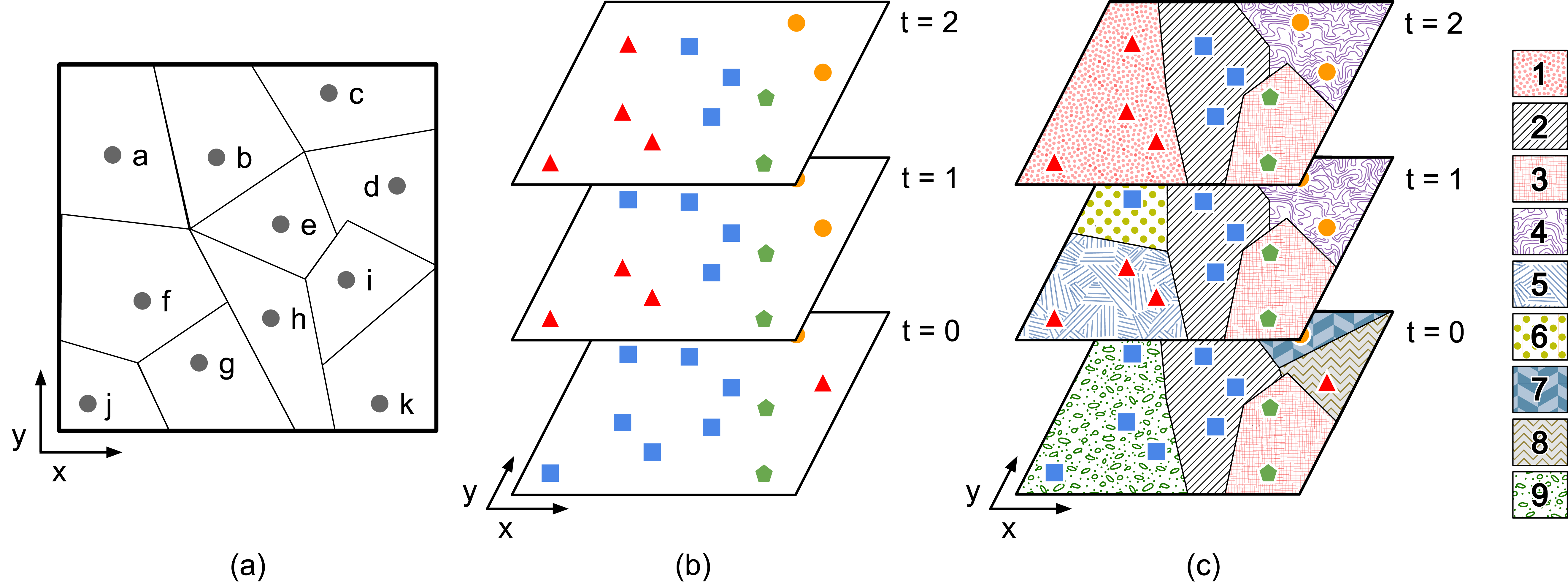}{}
\caption{(a) Sensors in a spatio-temporal dataset are shown at their spatial locations with the space decomposed into Voronoi polygons \cite{Aurenhammer:1991:VDS:116873.116880}. (b) Instances recorded at these sensors, coloured by their similarity. (c) The instances have been grouped into spatio-temporal regions, where each region is defined by the union of the Voronoi polygons of its constituent sensors, and a starting and ending time interval.}
\label{fig:regioning}
\end{figure}

To reduce the dataset $D$, we wish to find the set $R = \{r_{1},...,r_{|R|}\}$ of non-overlapping regions in the $T \times S^{\mathcal{D}}$ space. Each region $r_i \in R$ is defined by a bounding spatial polygon $P_i$ in the $S^{\mathcal{D}}$ space, a beginning time $t_{b}$ and ending time $t_{e}$. The subset of instances in $D$ that exist within the spatial and temporal bounds of region $r_i$ is denoted by $D_i$, where $D_i = \{d_{t,s} \in D \ |\  \mathit{ inside}(s,P_{i}), \ t_{b} \le t \le t_{e}\}$ and $\langle P_{i}, t_{b}, t_{e} \rangle = r_{i}$. Here, the function $\mathit{inside}(s,P_{i})$, is used to indicate that location $s$ is within the bounding polygon $P_{i}$ of region $r_i$.

The process of partitioning the $T \times S^{\mathcal{D}}$ space into regions, where $\mathcal{D} = 2$, is shown in Figure \ref{fig:regioning}. 
In (a), sensors from a dataset are shown at their locations in the 2-dimensional spatial domain, which has been split into Voronoi polygons surrounding each sensor.
In (b), instances recorded at the sensors have been colour coded by their similarity. 
For example, instances denoted by blue squares are more similar to each other than to instances denoted by red triangles, yellow circles and green pentagons. 
Finally, in (c), the instances from (b) have been grouped into 9 regions. Each region is defined by a spatial polygon and start and end time. 
For example, region 4 exists in the top-right of the space, from time 1 to time 2. 
This partitioning method is applicable to any number of spatial dimensions, however we focus on 2 spatial dimensions for simplicity.

Each region $r_i$ is associated with a model $m_j$ fitted to $D_i$, and we refer to the set of models as $M$. 
Furthermore, we denote $|m_{j}|$ to be the number of coefficients used to store model $m_{j}$. 
In $k$D-STR, we assume that a single modelling technique is used, thereby removing the need to store which modelling technique was used for each region.
We may form a model for each region, i.e., $|R| = |M|$, or associate a subset of regions with the same model, i.e., $|R| \neq |M|$.
Finally, the term \textit{reduction} is used to refer to the set of regions and models, denoted by the tuple $\langle R, M\rangle$.
The notation used in this paper is summarised in Table \ref{table:symbols}.

\begin{table}[t]
\centering
\caption{Notation used in this paper}
\label{table:symbols}
\begin{tabular}{p{2.5cm}p{10cm}}
\toprule
Symbol			& Definition \\ \hline
$D$				& Original dataset containing instances in the spatio-temporal space $T \times S^{\mathcal{D}}$, with the number of instances in $D$ denoted $|D|$ \\
$F$				& The set of real-valued features in $D$, excluding the referencing features $T$ and $S^{\mathcal{D}}$ \\
$d_{t,s}$		& An individual instance in $D$ \\
$d^{ \, f}_{t,s}$	& Value of $d_{t,s}$ for feature $f \in F$ \\
$R$				& A set of non-overlapping spatio-temporal regions in the $T \times S^{\mathcal{D}}$ space \\
$r_i$			& An individual spatio-temporal region in $R$ \\
$P_i$           & The set of $\mathcal{D}$-dimensional coordinates defining the bounding spatial polygon of region $r_i$ \\
$D_{i}$			& Set of instances of dataset $D$ contained in region $r_i$ \\
$M$				& Set of summary models of the instances in dataset $D$ \\ 
$m_{j}$			& Individual summary model, fitted over the a subset of instances in $D$, with the number of coefficients used to store $m_{j}$ denoted $|m_{j}|$ \\
$\langle R, M\rangle$				& A reduction of dataset $D$ \\
$e(D,\langle R, M\rangle)$	& Error introduced after $D$ is reduced to regions $R$ and models $M$ \\
$q(D,\langle R, M\rangle)$	& Ratio of storage required for regions $R$ and summary models $M$ compared to the original dataset $D$ \\
$h(D,\langle R, M\rangle)$	& Objective function used to find the best reduction given parameter $\alpha$, the constant that prioritises between $e(D,\langle R, M\rangle)$ and $q(D,\langle R, M\rangle)$\\
\bottomrule
\end{tabular}
\end{table}

When reducing $D$ to $\langle R, M\rangle$ we wish to minimise the information lost. 
One way of measuring information loss is to reconstruct the dataset from $\langle R, M\rangle$ as $D'$ and measure the difference between each instance $d_{t,s} \in D$ and the corresponding instance $d^{\prime}_{t,s} \in D^{\prime}$. 
A simple example is the mean absolute percentage error (MAPE) averaged across the dataset:

\begin{equation}
     e_{\text{MAPE}}(D,D') = \frac{1}{|D| \cdot |F|} \sum_{d_{t,s} \in D} \sum_{f \in F} 
		 \Bigg|\frac{d^{\, f}_{t,s} - d^{\prime \, f}_{t,s}}{d^{\, f}_{t,s}}\Bigg|
		 \label{eq:mape}
\end{equation}

However, MAPE is undefined for instances containing feature values of 0, such as rainfall datasets. 
Therefore, it is unsuitable as a measurement of error in the reduction of such datasets. An alternative measure is the normalised root mean square error (NRMSE) averaged across the dataset:

\begin{equation}
	e_{\text{NRMSE}}(D,D') = \frac{1}{|F|} \sum_{f \in F} \frac{\psi(f, D, D')}{\mathit{range}(f)} 
	\label{eq:nrmse}
\end{equation}
where,

\begin{equation}
  \psi(f, D, D') = \sqrt{\frac{\sum_{d_{t,s} \in D} (d^{\, f}_{t,s} - d^{\prime \, f}_{t,s})^2}{|D|}}
\end{equation}
and $\mathit{range}(f) = \mathit{max}_{t,s}(d^{\, f}_{t,s}) - \mathit{min}_{t,s}(d^{\, f}_{t,s})$. In general, we refer to the error introduced by reducing $D$ to $\langle R, M\rangle$ as $e(D,\langle R, M\rangle)$. When using a difference metric between $D$ and the reconstructed $D'$, we say $e(D,\langle R, M\rangle) = e(D,D')$.

Whilst minimising the error introduced in reduction, we also wish to minimise the storage cost of the reduced dataset. 
To store an instance in the original dataset, a location point for each of the $\mathcal{D}$ spatial dimensions, a time in the temporal dimension, and a value for each of the $F$ real-valued features must be stored.
The storage cost of the original dataset is then the cost per instance multiplied by the number of instances: 

\begin{equation}
	\label{eq:storageD}
	\mathit{storage}(D) = |D| \cdot (|F| + k)
\end{equation}

In the case of the reduced dataset, the storage required for each region is a start and end time in the temporal dimension, the set $P_{i}$ of points that define the bounding polygon of the region in the spatial domain, and the set $m_j$ of coefficients required to store the model of the region. 
Each point in the set $P_{i}$ requires a value for each of the $\mathcal{D}$ spatial dimensions to be stored. 
Therefore, the storage cost of the reduced dataset is the sum of region costs as shown in Equation \ref{eq:storageR}, where $(|P_{i}| \cdot (k-1))$ is the number of values stored for the spatial boundary of a region $r_i$, and 2 values are used to store the beginning and end times of $r_i$. Furthermore, $|m_{j}|$ is the number of coefficients used to store the model $m_j$.

\begin{equation}
	\label{eq:storageR}
	\mathit{storage}(\langle R, M\rangle) = \sum_{i=1}^{|R|} \big((|P_{i}| \cdot (k-1)) + 2 \big) + \sum_{j=1}^{|M|} |m_{j}|
\end{equation}

We use the ratio of the storage cost of the reduced and original datasets to define the storage ratio, as shown in Equation \ref{eq:storage}.

\begin{equation}
	\label{eq:storage}
     q(D,\langle R, M\rangle) = \frac{\mathit{storage}(\langle R, M\rangle)}{\mathit{storage}(D)}
\end{equation}

Finally, in reducing a dataset we wish to balance the reduction in storage against the error introduced. 
The best balance is subjective and depends on the dataset being reduced as well as the kinds of analysis to be performed after reduction. 
Given the user's preference, we can use the objective function $h(D,\langle R, M\rangle)$ defined in Equation \ref{eq:objective-function} to find the optimal reduction. 
Here, the parameter $\alpha$ is used to indicate the user's preference for the balance between reduction in storage and minimising the error introduced. 
The parameter $\alpha$ is bounded to the range $[0,1]$ and must be determined before reducing the dataset.

\begin{equation}
	\label{eq:objective-function}
	h(D,\langle R, M\rangle) = \alpha \cdot q(D,\langle R, M\rangle) \: + \: (1-\alpha) \cdot e(D,\langle R, M\rangle)
\end{equation}

\section{\lowercase{\textit{k}}D-STR: \lowercase{\textit{k}}-Dimensional Spatio-Temporal Reduction}
\label{sec:methodology}
In this paper, we introduce the $k$-Dimensional Spatio-Temporal Reduction method ($k$D-STR) for achieving the goal of reduction described in Section \ref{sec:preliminaries}. 
As input, the method takes the dataset to be reduced, the parameter $\alpha$, which indicates the user's preference for reduction in storage volume versus introduced error, and a preferred modelling technique. 
$k$D-STR is an iterative algorithm that begins by forming a partitioning tree over the dataset. 
Then, starting with a single region at the root of the tree, a model is fitted to the instances within the region to summarise the data. 
Next, $k$D-STR iteratively determines whether to partition the $T \times S^{\mathcal{D}}$ space into more non-overlapping regions, or to increase the model complexity of one of the existing regions, with the aim of improving its accuracy. 
The decision taken at each step is that which minimises the objective function $h(D, \langle R, M\rangle)$. 
In this section, we describe each of the steps of $k$D-STR.

\subsection{Data Partitioning}
\label{sec:data-partitioning}
By identifying spatio-temporal regions of similar instances, $k$D-STR is able to reduce a dataset to a set of regions and models. Whilst methods exist for creating a partition tree over a dataset, such as quadtree and octree decomposition \cite{Samet:1984:QRH:356924.356930}, the number of new partitions introduced by these methods at each level of partitioning is fixed. Instead, it is more beneficial to use the variation of feature values over space and time to determine how many partitions are introduced at each level of the partition tree. $k$D-STR uses the difference between instances within the data itself to determine how many regions should be placed over the $T \times S^{\mathcal{D}}$ space at each level of partitioning. Each region is defined by a piece-wise linear polygon in the spatial domain and a beginning and ending time in the temporal domain.

\begin{figure}[t]
     \centering
          \includegraphics[width=\textwidth]{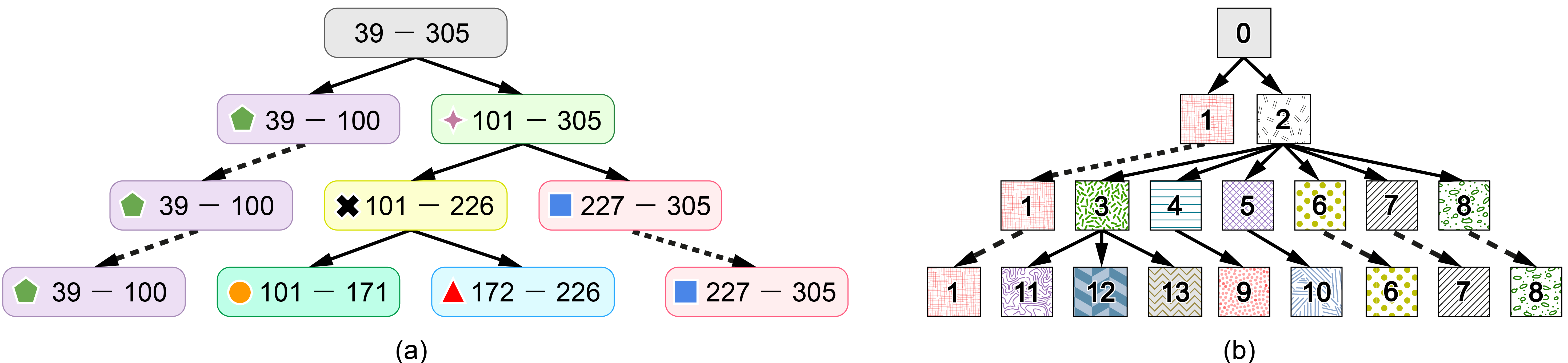}%
     
     \caption{(a) The cluster tree created by hierarchical clustering on the data in Table \ref{table:footfall}. Each node in the cluster tree shows the bounds defining the cluster. Solid arrows are used to show which clusters decompose into new clusters at each level of the tree, and dashed arrows are used to show clusters that remain the same. (b) The resulting partitioning tree. }
     \label{fig:clusters}
\end{figure}

To begin partitioning the dataset, the instances are first clustered into a set of clusters $C$ using hierarchical agglomerative clustering in the feature space. 
By clustering instances in the feature space, rather than clustering in the $T \times S^{\mathcal{D}}$ space, $k$D-STR clusters together instances that have similar feature values regardless of when and where the instances were recorded. 
Hierarchical clustering is used as the clusters found in the feature space are not expected to be globular and the resultant \textit{cluster tree} allows regions and models to be retained in some areas or time periods as the number of clusters required changes \cite{6832486}.
To illustrate this further, consider the dataset shown in Table \ref{table:footfall} containing data from footfall sensors that record the number of people that walk through an area at three consecutive time steps. 
The data is hierarchically clustered using just the raw footfall count values. This results in the cluster tree shown in Figure \ref{fig:clusters}(a), where each node in the tree shows the boundaries of the clusters in the footfall feature space. 

\begin{table}[t]
     \centering
     \caption{Example footfall data recorded at the 9 sensors (A-K) shown in Figure \ref{fig:regioning}.}
      \label{table:footfall}
     \begin{tabular}{@{}lllllllllllll@{}}
          \toprule
           &  & \multicolumn{11}{c}{\textbf{Sensor}} \\
           &  & \textbf{A} & \textbf{B} & \textbf{C} & \textbf{D} & \textbf{E} & \textbf{F} & \textbf{G} & \textbf{H} & \textbf{I} & \textbf{J} & \textbf{K} \\ \midrule
          \multirow{3}{*}{\textbf{Time step}} & \textbf{0} & \cSquare252 & \cSquare278 & \cCircle148 & \cTriangle193 & \cSquare279 & \cSquare248 & \cSquare267 & \cSquare296 & \cHexagon45 & \cSquare241 & \cHexagon58 \\
           & \textbf{1} & \cSquare247 & \cSquare305 & \cCircle153 & \cCircle145 & \cSquare301 & \cTriangle212 & \cTriangle207 & \cSquare292 & \cHexagon67 & \cTriangle201 & \cHexagon52 \\
           & \textbf{2} & \cTriangle210 & \cSquare296 & \cCircle139 & \cCircle134 & \cSquare299 & \cTriangle199 & \cTriangle192 & \cSquare287 & \cHexagon39 & \cTriangle189 & \cHexagon46 \\ \bottomrule
          \end{tabular}
 \end{table}

After the cluster tree is formed, regions are found for each level. 
For a given level of the cluster tree, each instance in the $T \times S^{\mathcal{D}}$ space is labelled with the cluster it has been grouped into. 
Then homogeneous regions that are connected components belonging to the same cluster in the $T \times S^{\mathcal{D}}$ space are found. 
Since storing the bounding spatio-temporal polygon of each homogeneous region may require many coordinates to be stored, we assert that each region must be defined by a single start and end time to limit the shapes in the $T \times S^{\mathcal{D}}$ space that regions may take.

\begin{figure}[t]
     \centering
          \includegraphics[width=0.8\textwidth]{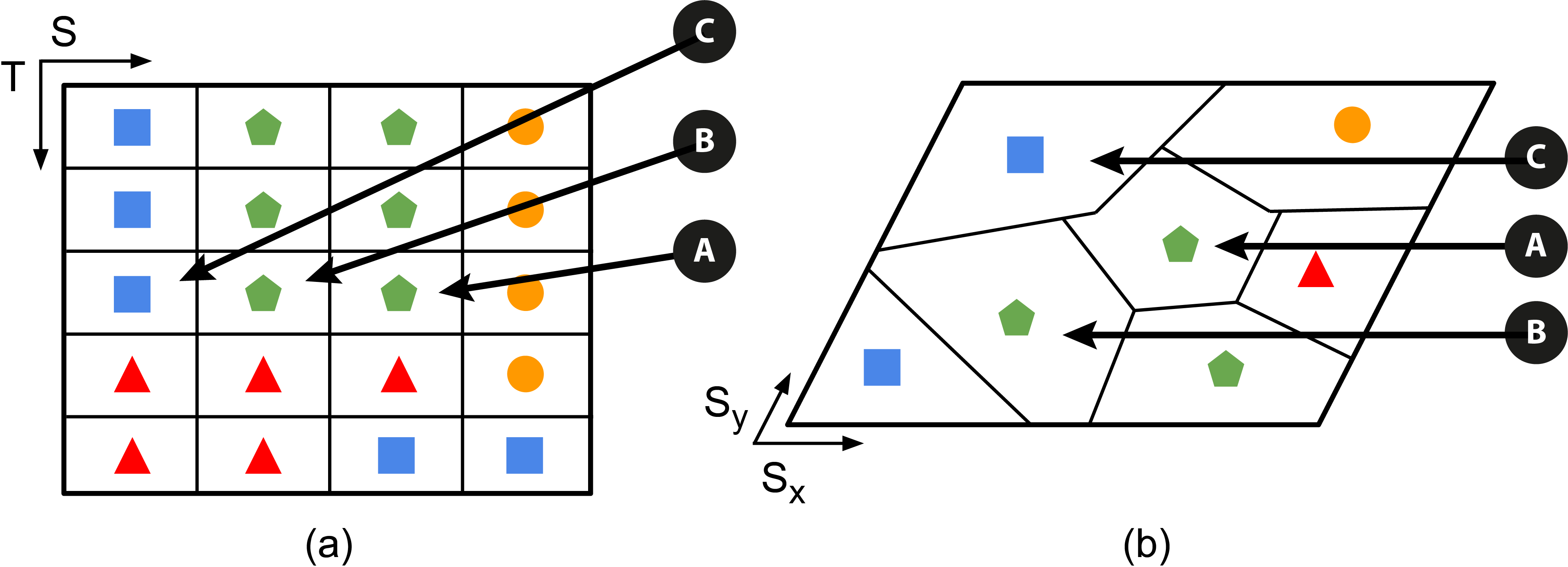}%
     
     \caption{(a) Instance \textbf{A} is chosen as the start of a new region. The region is expanded left to include instance \textbf{B} but not instance \textbf{C} as it belongs to a different cluster. (b) Instance \textbf{A} is chosen as the start of a new region. However, the region cannot be expanded left as both instances \textbf{B} and \textbf{C} are neighbours on the left of instance \textbf{A}.}
     \label{fig:kd-region-growing}
\end{figure}

However, finding connected components that are limited in shape is not trivial.
In previous work, 2D-STR found a connected component in 2 dimensional space (i.e., 1 spatial dimension and 1 temporal dimension) by selecting an instance at random as the start of a new region, and growing the region as far as possible in each direction of each dimension \cite{2d-str}. 
That is, when the dataset is viewed as a 2-dimensional matrix, a region could be expanded beyond a beginning instance by moving as far left and right as possible, then as far up and down as possible.
In the example shown in Figure \ref{fig:kd-region-growing}(a), if instance \textbf{A} is chosen as the beginning instance of a new region, the region is first grown by extending the left boundary leftwards, adding instance \textbf{B} to the region, and then stopping as instance \textbf{C} belongs to a different cluster and cannot be added to the region.  
This process is only possible as there is a natural ordering to the sensors in the spatial and temporal dimensions.
For the spatial dimension, sensors are naturally ordered by their position along the dimension.
However, when the number of spatial dimensions is greater than 1, this natural ordering disappears and so it is ambiguous how to expand each region in the spatial domain.
In the 3 dimensional example shown in Figure \ref{fig:kd-region-growing}(b), suppose that instance \textbf{A} is again chosen as the start of a new region. If the region was expanded left in the $S_x$ direction, both instances \textbf{B} and \textbf{C} would be added to the region. Adding instance \textbf{B} is acceptable but adding instance \textbf{C} would break the region's homogeniety. 
Therefore, the aim is to add instance \textbf{B} to the region without having to add instance \textbf{C}.

To overcome this issue, we first partition the spatial domain $S$ into discrete polygons around each sensor using Voronoi partitioning \cite{Aurenhammer:1991:VDS:116873.116880} as shown in Figure \ref{fig:regioning}(a). 
A similar action is also performed for the temporal domain $T$, forming a discrete timestep around each unique time present in $D$. 
By discretising the spatio-temporal dimensions, we can view the instances in $D$ as a spatio-temporal graph, where each vertex is an instance and edges link vertices that are \textit{adjacent}. We say two instances are adjacent if:

\begin{enumerate}[label=(\roman*)]
    \item they were recorded consecutively at the same sensor, or
    \item they were recorded at the same time and the polygons surrounding their sensors are adjacent in the discretised spatial domain.
\end{enumerate}

After discretising the spatial and temporal domains, $k$D-STR extends regions in a breadth-first manner. 
After an instance is chosen as the start of a new region, all instances that are adjacent to this instance in the spatial domain and belong to the same cluster are added to the region. 
In Figure \ref{fig:kd-region-growing}(b), this would add instance \textbf{B} to the region but not instance \textbf{C}. 
Then, after all spatial neighbours of the initial instance are considered, the temporal boundary is extended by up to 1 timestep before and after the initial instance if doing so does not break the cluster homogeneity of the region. 
This process is repeated continuously, expanding the boundaries of the region by a depth of 1 neighbour to the existing instances in the region spatially, and 1 timestep before and after the region, until the spatial and temporal bounds of the region cannot be expanded any further.

\begin{figure}[t]
     \centering
          \includegraphics[width=\textwidth]{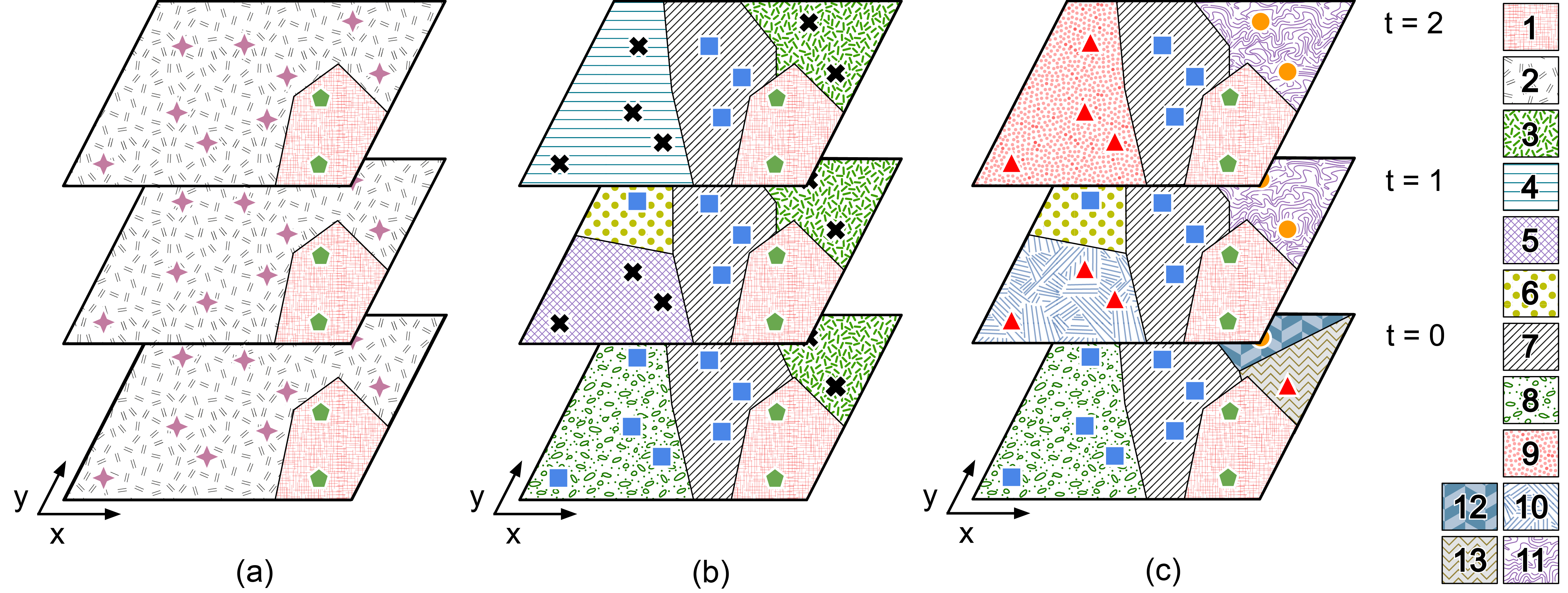}%
     
     \caption{The spatio-temporal regions formed by the data partitioning process, when applied to the data shown in Table \ref{table:footfall}. Subfigures (a), (b) and (c) show the partitions at levels 2, 3 and 4 of the partitioning tree respectively. Subfigure (c) is also shown in Figure \ref{fig:regioning}(c), and regions 1--9 in Figure \ref{fig:regioning}(c) correspond to regions 1 and 6--13 here.}
     \label{fig:regions}
\end{figure}

Converting a level of the cluster tree into a level of the partition tree is complete when all instances in $D$ are associated with a homogeneous region within that level of the partition tree. 
The result is a partition tree in which each instance is assigned to a region at each level of the tree. 
The relationship between clusters and regions is shown in Figure \ref{fig:clusters}. 
The first four levels of the cluster tree are shown in subfigure (a), and the corresponding first four levels of the partitioning tree are shown in subfigure (b). 
Each level of the partitioning tree is also shown in Figure \ref{fig:regions}. 
At the root of the cluster tree, at level 1 with 1 cluster, all 33 instances shown in Table \ref{table:footfall} are placed into a single region, namely region 0. 
On level 2 of the cluster tree, with 2 clusters, region 0 is decomposed into regions 1 and 2 in the partitioning tree. 
On level 3, with 3 clusters, the \textit{101 -- 305} cluster has been decomposed into 2 more clusters and region 2 has been decomposed into regions 3 to 8 accordingly. 
Finally, on level 4 of the cluster tree, the \textit{101 -- 226} cluster has been decomposed and region 3 decomposed into regions 11--13, region 4 into region 9 and region 5 into region 10.
Whilst the sensors and time periods covered by regions 4 and 5 do not change, the parent cluster was decomposed and so they were replaced by regions 9 and 10 respectively.

By using hierarchical clustering, we ensure that only some regions in the partitioning tree are decomposed between levels. 
This results in some regions remaining unchanged throughout multiple levels of the partitioning tree, and this can be exploited in the reduction process by retaining models of these regions during reduction. 
Furthermore, the models of regions that are replaced but retain the same sensors and time period can also be retained.

\subsection{Region Modelling}
\label{sec:region-modelling}
After partitioning $D$ into a hierarchy of regions, a technique is required to model the instances within each region. 
We may form a model for the instances within each region or, since each region is associated with a cluster, we may form a model per cluster and link all regions belonging to the same cluster to the same model.
To maximise the utility of the reduction, we require the ability to reconstruct the instances of the original dataset from the models output. 
Furthermore, we wish to enable the imputing of instances in spatial and temporal locations that have not been sampled in the dataset, but have nearby sensors or time periods. 
If the type of analysis to be performed after reduction is already known and the task of imputation is not necessary, more appropriate modelling techniques may be used that offer higher accuracy, smaller storage or permit faster answering of queries. Overall, we wish the storage requirements of each model to be less than the storage requirement of the original instances the model was fitted to.

For each region $r_i$ or cluster $c_i$ in each level of the partitioning tree, a model $m_j$ is fitted to the instances within the region or cluster. 
The spatial and temporal values of the instances are used as the independent or predictor values of the model, whilst the feature values of the instances are used as the dependent or response values of the model.

In this paper, we consider three illustrative modelling techniques. 
First, we consider polynomial linear regression (PLR) owing to its ability to explain data that is spatially and temporally autocorrelated. 
Polynomial regression is a fundamental technique in statistical modelling that is both easy to interpret and permits interpolation. 
Second, we consider discrete cosine transform (DCT) approximation for its ability to model periodic data using few coefficients. 
By removing low-weighted coefficients and storing just those weighted highly, the original data can be reproduced with reasonable accuracy. 
Finally, we consider decision tree regression (DTR) owing to the interpretability of the models output. 
For each of these techniques, we may form a model per region or a model per cluster, and so we refer to these as PLR-R and PLR-C, DCT-R and DCT-C, and DTR-R and DTR-C respectively.

\subsection{Data Reduction}
The $k$D-STR algorithm can be seen in Algorithm \ref{algo:kd-str}, in which a model is formed for each region, though the algorithm may be easily adapted to model on clusters instead. Prior to initialising $k$D-STR, a value for the parameter $\alpha$ must be chosen, where $0 \leq \alpha \leq 1$. This parameter weights the error introduced against the storage ratio in the objective function (Equation \ref{eq:objective-function}). Since a value of 0 indicates a preference for minimising the error introduced with no consideration of the storage used, $k$D-STR would not stop iterating until the error metric $e(D,\langle R, M\rangle) = 0$. Such a perfect model may be unrealistic and less efficient than storing the original dataset. Similarly, when a value of 1 is chosen, both increasing the complexity of a model in $M$ and decomposing the spatio-temporal space into more regions would increase the quantity of storage used beyond that of the initialisation step. Therefore, $k$D-STR would not iterate beyond the first iteration. Both of these scenarios may not be useful and so values of 0 or 1 should likely be avoided for the parameter $\alpha$.

\begin{algorithm}
     \caption{The algorithm $k$D-STR}
     \label{algo:kd-str}

     clusterTree = cluster($D$)\;
     numberClusters = 1\;
     $R$ = findRegions($D$, clusterTree, numberClusters)\;
     $M$ = \{\} \tcp*{Initialise the set of models to the empty set}
     \For{$r_i$ \textbf{in} $R$}{
          $m_i$ = model($D_i$, 1) \tcp*{Model the data in region $r_i$ using the simplest complexity}
          $M$.add($m_i$)\;
     }

     $h$ = heuristic($D, R, M$) \tcp*{Calculate heuristic for 1 region and simple model}

     \tcp{Now iterate until heuristic $h$ is minimised}
     \Do{$h_1$ < h \textbf{or} $h_2$ < h}{
          \tcp{First, check if increasing an existing model's complexity minimises $h$ further}
          $h_{1} = h$\;
          \For{$r_i$ \textbf{in} $R$}{
               $M'$ = $M$\;
               $m'_{i}$ = model($D_i$, $m_{i}$.complexity + 1)\;
               $M'$.replace($m_i, m'_{i}$) \tcp*{Replace $m_i$ with $m'_{i}$ in $M'$}

               $h'$ = heuristic($D, R, M'$)\;
               \If{$h' < h_{1}$}{
                    $h_{1} = h'$\;
                    $M_{best} = M'$\;
               }
          }

          \tcp{Second, check if increasing the number of regions minimises $h$ further}
          $R'$ = findRegions($D$, clusterTree, numberClusters+1)\;
          $M''$ = \{\}\tcp*{Initialise the set of models to the empty set}

          \For{$r_i$ \textbf{in} $R'$}{
               \If{$r_i$ \textbf{in} $R$}{
                    $M''$.add($m_i$) \tcp*{Add $m_i \in M$ to $M''$}
               }
               \Else{
                    $m''_{i}$ = model($D_i$, 1)\;
                    $M''$.add($m''_{i}$)\;
               }
          }
          $h_2$ = heuristic($D, R', M''$)\;

          \tcp{Finally, if increasing the number of regions, or the complexity of an existing model, is more optimal than $h$, take that choice}
          \If{$h_1 < h_2$ \textbf{and} $h_1 < h$}{
               $M = M_{\text{best}}$\;
               $h = h_1$\;
          }
          \ElseIf{$h_2 < h_1$ \textbf{and} $h_2 < h$}{
               $R = R'$\;
               $M = M''$\;
               $h = h_2$\;
               numberClusters = numberClusters + 1\;
          }
     }
     \Return{$R$, $M$}
\end{algorithm}

After a partitioning tree has been formed for the dataset $D$ (Algorithm \ref{algo:kd-str}, lines 1--3), and a value for $\alpha$ and a modelling technique have been chosen, the reduction steps of $k$D-STR are performed. 
$k$D-STR begins at the root of the partitioning tree, with all instances belonging to a single region. 
This region is then modelled using the simplest form of the modelling technique used (lines 5--7): in the case of polynomial regression a polynomial model of order 0 (simply a mean) is constructed for each feature; in the case of DCT only the highest weighted cosine coefficient is considered; and, in the case of DTR the decision tree is limited to a depth of 1. Note that the value 1 is used in line 6 to indicate the simplest form of model.
After the model is fitted to the data, the result of the objective function $h(D, \langle R, M\rangle)$ is calculated for this first reduction step (line 8).

After the first reduction step, the algorithm iterates. On each iteration, $k$D-STR decides whether to increase the complexity of one of the existing models (lines 11--18) or increase the number of regions in the partitioning of the $T \times S^{\mathcal{D}}$ space (lines 19--27). When the number of regions is increased, only some existing regions are decomposed. For regions that are not decomposed, the models for these regions persist thereby improving the efficiency of $k$D-STR (lines 22--23). For previous regions that are decomposed, new regions are found and new models are fitted to these regions (lines 24--26).

The $k$D-STR algorithm stops when the objective function cannot be minimised further, i.e., $h_1 \ge h(D, \langle R, M\rangle)$ and $h_2 \ge h(D, \langle R, M\rangle)$. When the algorithm stops the reduction is complete and the set of regions and models, $\langle R, M\rangle$, is returned.

\subsection{Analysis of Running Time and Memory}
To understand the time and memory complexities of $k$D-STR, we consider the startup cost of clustering the dataset and the cost of each iteration of the algorithm separately. 
In the startup phase of the algorithm (line 1 of Algorithm \ref{algo:kd-str}), hierarchical clustering is performed. 
Whilst hierarchical agglomerative clustering runs in $\mathcal{O}(|D|^3)$ time and requires $\mathcal{O}(|D|^2)$ memory, a more efficient approximation has been demonstrated to reduce the time complexity to $\mathcal{O}(|D|^2)$ \cite{fastcluster}. 
Therefore, the startup phase of $k$D-STR requires $\mathcal{O}(|D|^2)$ time and $\mathcal{O}(|D|^2)$ space.

After clustering, the algorithm iterates in successive rounds, either increasing the number of regions in the reduced dataset or increasing the complexity of a model. 
To increase the number of regions, the number of clusters is increased by 1. Each instance is labelled with the ID of its cluster in the next level of the cluster tree, and homogeneous regions are found for each cluster at this next level.
To create a region, an instance that has not been assigned to a region is chosen at random as the start of the next region, and other instances adjacent to the instance are added to the region if they belong to the same cluster. 
This process is repeated in rounds until no more instances can be added to the region.
During this process, the boundary between any two adjacent instances is checked at most twice. 
First, the boundary is checked in one direction to see if the two instances belong to the same cluster. 
If they do not, the second instance is not added to the region.
In this case, at a later time of processing the second instance will be added to a different region and the same boundary is checked again to see if the first instance can be added to the new region.
Since that instance already belongs to a region it will not, but the adjacency edge between the two instances has now been considered twice.
This process requires $\mathcal{O}(x|D|)$ time, where $x$ is the maxmimum number of instances that are adjacent to any single instance in $D$. 
Furthermore, this process requires $\mathcal{O}(|D|)$ memory to store the cluster and partition ID of each instance, as well as a working list of the boundary instances of the region being expanded.

As well as increasing the number of regions on each iteration, the algorithm also considers increasing the complexity of an existing model. 
For example, in the case of polynomial linear regression (PLR), this requires $\mathcal{O}(y^2|D|)$ time and $\mathcal{O}(y^2)$ memory per model, where $y$ is the number of coefficients calculated for the model. Discrete cosine transforms (DCT) can be performed in $\mathcal{O}(|D|^2)$ time and $\mathcal{O}(1)$ memory per model, although fast cosine transforms can be performed in $\mathcal{O}(|D| \log |D|)$ time. 
Similarly, decision tree regression (DTR) can be performed in $\mathcal{O}(k|D|^2)$ time and $\mathcal{O}(k)$ memory per model. 
Therefore, the startup phase of $k$D-STR requires $\mathcal{O}(|D|^2)$ time and $\mathcal{O}(|D|^2)$ memory, whilst each iteration, in the case of PLR, requires $\mathcal{O}((x + y^2|M|)|D|)$ time and $\mathcal{O}(|D| + y^2|M|)$ memory.

% DTR
% We have to decide what the best break point is in each dimension, i.e. k|D|. If the tree was balanced we'd then have to do that log |D| times, but it might not be. Thus it can be  

\section{Evaluation Methodology}
\label{sec:evaluation}
To evaluate $k$D-STR, we compared the rate at which the data size decreased with the rate of increased error for three spatio-temporal datasets taken from different sources. 
To measure the error introduced by the reduction process, we considered the normalised root mean square error (NRMSE) as shown in Equation \ref{eq:nrmse}. 
The NRMSE metric was used as it indicated how well each feature is modelled by the reduction process as a percentage error of the range of original values. % This indicates how well we can reconstruct the low-frequency changes of the feature.
To measure the reduction in storage between the original and the reduced datasets, we used the storage ratio defined in Equation \ref{eq:storage}. 
This indicated how much smaller the reduced dataset was compared to the original dataset. 
The number of regions in the reduced dataset was also considered, since this indicated how many data points in the reduced dataset have to be processed for many types of query and analysis.

To evaluate the generality of $k$D-STR to a range of spatio-temporal data, we considered datasets that exhibited different distributions and characteristics in space and time.
We considered three sources of data: air temperature sensors, traffic counting sensors and rainfall sensors. 
Each of these datasets exhibited different spatial and temporal variances, that is the dissimilarity between nearby instances in space and time. 
Furthermore, whilst the air temperature and road traffic datasets continuously evolved, meaning there was a predictable trend of increasing and decreasing feature values over space and time, the rainfall dataset was event driven.
Rainfall events were localised to small groups of sensors at the same time, and there were discontinuities between no rainfall (a recording of 0mm rainfall) and rainfall (non-zero values) beween consecutive time intervals at the same sensor.
We expected the reduction of datasets exhibiting higher variance in space to yield more regions in space than time, and vice versa.    
The characteristics of the three datasets used are shown in Table \ref{table:dataset-characteristics}. 
Using these characteristics, other datasets may be likened to those tested.
The spatio-temporal datasets used to evaluate $k$D-STR in this paper are as follows.

% Please add the following required packages to your document preamble:
% \usepackage{booktabs}
\begin{table}[]
\caption{Characteristics of the datasets used for evaluation}
\label{table:dataset-characteristics}
\begin{tabular}{p{1.9cm}p{1.4cm}p{1.4cm}p{4.1cm}p{3.1cm}}
\toprule
 & Spatial Variance & Temporal Variance & Discontinuities in Space & Discontinuities in Time \\ \midrule
Air \mbox{Temperature} & Low & Low & Low, nearby sensors recorded similar feature values & Low, small variations between adjacent time intervals \\
Traffic & Low & High & Nearby sensors on main carriageway recorded similar feature values at same time, but differed substantially from sensors on slip roads; road traffic collisions introduced discontinuities in space & Road traffic collisions introduced discontinuities in time \\
Rainfall & Changed over time & Low & Rainfall events are often localised to groups of nearby sensors at the same time & Varied substantially over time as rainfall events occured \\ \bottomrule
\end{tabular}
\end{table}

\begin{enumerate}
    \item \textbf{Air Temperature}: 12 samples of month-long air temperature data were collected from the Met Office Integrated Data Archive System (MIDAS) Land and Marine Surface Stations Dataset in the United Kingdom (UK) \cite{metofficeweather}. The samples used were taken from January to December 2017 and covered all Met Office temperature recording sensors in the UK, which recorded one instance per hour. The dataset contained three real-valued features: \textit{temperature ( $^{\circ}$\!C)}, \textit{wet bulb temperature ( $^{\circ}\!$C)} and \textit{dew point ( $^{\circ}$\!C)}. The values of these features fluctuated throughout the day but followed a daily trend of increasing and decreasing values. The samples used from this dataset contained between 240,201 and 266,197 instances each. 
    
    \item \textbf{Traffic}: A set of 28 samples of the WebTRIS traffic dataset was evaluated \cite{webtris}. Each sample was a month-long survey of traffic counting sensors taken from several roads in England over April 2017, September 2017, November 2017 and December 2017. These samples were taken from the A30, A66 and A69 trunk roads and the M1, M11, M20, and M56 motorways. These roads were chosen for their differing spatial distributions and traffic characteristics. The traffic dataset contained six real-valued features: \textit{count of vehicles of length 0m to 5.2m}, \textit{count of vehicles of length 5.21m to 6.6m}, \textit{count of vehicles of length 6.61m to 11.6m}, \textit{count of vehicles of length 11.61m or greater}, \textit{total count of vehicles} and \textit{average speed (mph)}. Similar to the air temperature dataset, the traffic dataset also exhibited daily trends as well as weekly trends. Sensors on slip roads (entries and exits) were interspersed amongst main carriageway sensors, and exhibited much lower traffic counts compared to the main carriageway sensors. The samples used from this dataset contained between 54,180 and 86,042 instances each. 
    
    \item \textbf{Rainfall}: 12 samples of month-long rainfall data were used from the Met Office Integrated Data Archive System (MIDAS) \cite{metofficeweather}. Again, the samples used were taken from January to December 2017 and covered all Met Office rainfall recording sensors in the UK at hourly intervals\footnote{Note the stations that recorded rainfall were not the same as the stations that recorded air temperature, and so this dataset contained a different spatial distribution to the air temperature dataset.}. The dataset contained a single feature, \textit{precipitation (mm)}. However, unlike the air temperature and traffic datasets there was not a temporal pattern that was expected to be observed every day. Instead, the dataset contained many instances of 0mm rainfall, especially in the summer months. It was expected this property would make the dataset more efficient to reduce than the traffic and temperature datasets, since large spatio-temporal regions could be reduced to a mean value of 0mm. The samples used from this dataset contained between 194,371 and 215,119 instances each.
\end{enumerate}

We used 6 modelling techniques in our evaluation of $k$D-STR: polynomial linear regression modelling on each region (PLR-R), polynomial linear regression modelling on each cluster (PLR-C), discrete cosine modelling on each region (DCT-R), discrete cosine modelling on each cluster (DCT-C), decision tree regression on each region (DTR-R) and decision tree regression on each cluster (DTR-C). We evaluated these methods on both the clusters and regions generated by the data partitioning stage to test the hypothesis that modelling on clusters would output few but complex models and modelling on regions would output many simple models.

Alongside the 6 modelling techniques tested on the presented datasets, 5 values of $\alpha$ were tested, namely, $\alpha \in \{0.1, 0.25, 0.5, 0.75, 0.9\}$. Finally, to compare $k$D-STR with other reduction methods for spatio-temporal datasets, we also used IDEALEM \cite{8085035}, DEFLATE \cite{deutsch1996deflate} and PCA adapted for spatio-temporal sensor \cite{pearson1901liii, demvsar2013principal}.

\section{Results and Discussion}
\label{sec:results}
In this section, we discuss the results of applying $k$D-STR to the datasets discussed in Section \ref{sec:evaluation}. 
We compare the results of the different modelling techniques and the effects of the parameter $\alpha$. 
Furthermore, we discuss reduction when multiple spatial referencing systems (SRSs) can be used, as well as the spatial and temporal properties of the datasets that can be found using the partitioning of space and time by $k$D-STR.

\subsection{The Trade-off Between Error and Storage}
\label{sec:results-alpha}
\begin{sidewaysfigure}
     \centering
     \vspace{7cm}
     \subfloat[Results of $k$D-STR on the air temperature dataset (12 1-month samples)]{
          \includegraphics[width=0.87\textheight]{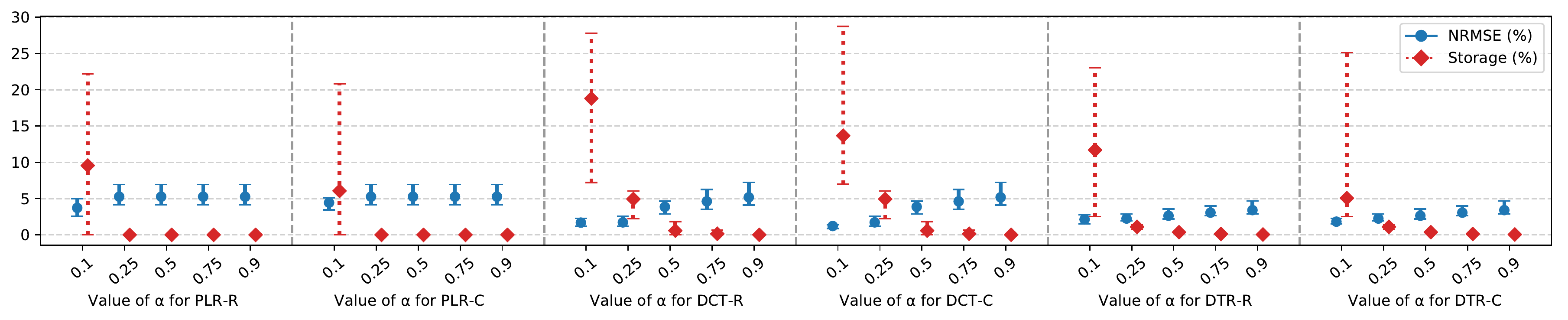}%
     }

     \subfloat[Results of $k$D-STR on the traffic dataset (28 samples, collected over 4 months from 7 roads)]{
          \includegraphics[width=0.87\textheight]{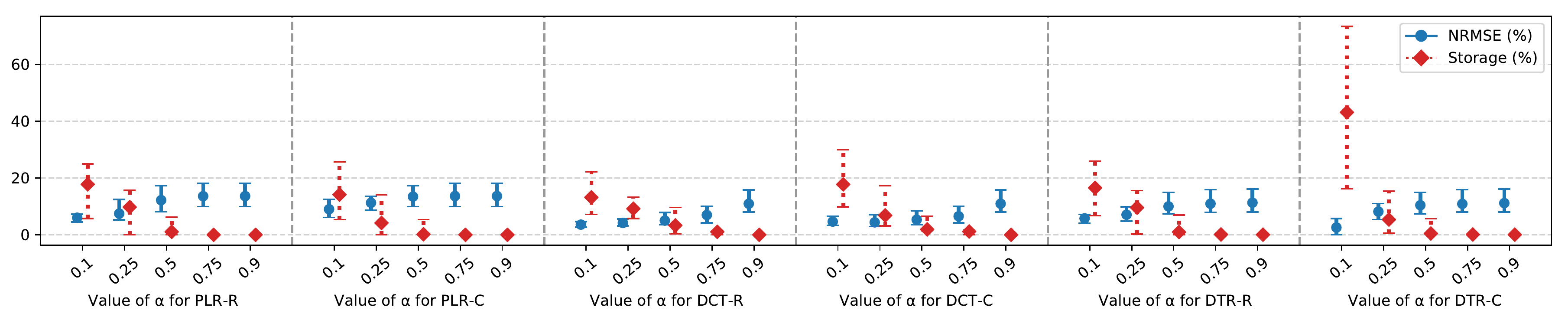}%
     }

     \subfloat[Results of $k$D-STR on the rainfall dataset (12 1-month samples)]{
          \includegraphics[width=0.87\textheight]{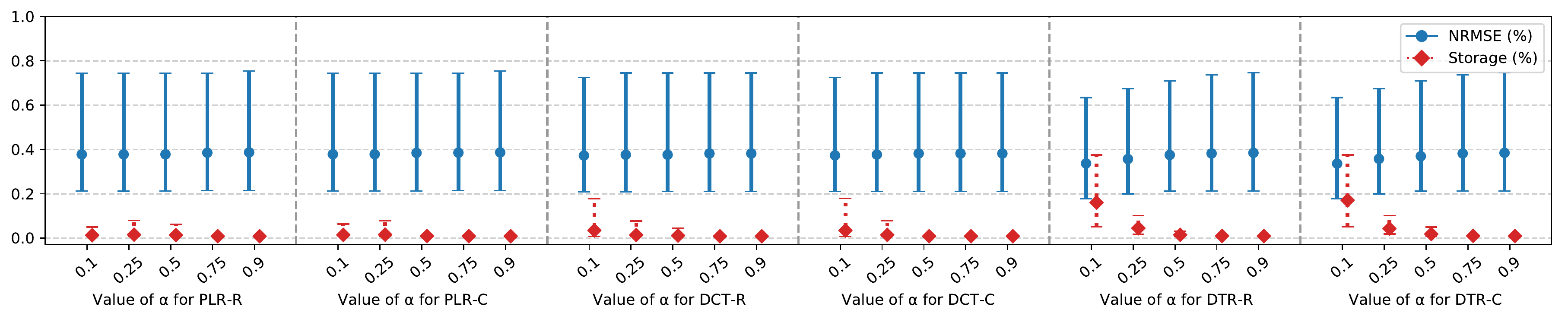}%
     }
     
     \caption{Error introduced and storage used by $k$D-STR on the air temperature, traffic and rainfall datasets. Each sub-figure shows the results of $k$D-STR using the six modelling techniques evaluated for varying values of $\alpha$.}
   \label{fig:nrmse-storage}
   \vspace{-7cm}
\end{sidewaysfigure}

For each of the three datasets evaluated, the mean and range of the NRMSE and storage ratios of the $k$D-STR reduction are shown in Figure \ref{fig:nrmse-storage}. 
A subfigure is shown per dataset, with each showing the results of using each of the 6 modelling techniques discussed in Section \ref{sec:evaluation}. Furthermore, each subfigure shows the results of the reduction process given the five values for the parameter $\alpha$ described in Section \ref{sec:evaluation}.

Several conclusions can be drawn from these results. In all cases, the error introduced by the reduction was smaller for lower values of $\alpha$ than for higher values. Conversely, the storage ratio of data output was greater for lower values of $\alpha$ than for higher values. For DCT-R modelling on the air temperature dataset, the storage quantity used in the reduced datasets ranged from 7.2\% to 27.8\% when $\alpha = 0.1$, but decreased to 0.006\% to 0.1\% when $\alpha=0.9$. Conversely, the error introduced ranged from 1.2\% to 2.3\% when $\alpha=0.1$ but increased to 4.1\% to 7.2\% when $\alpha=0.9$. A similar relationship between the error introduced and storage quantity used was observed across all of the datasets presented.

In some cases, particularly when modelling with polynomial regression, this trend plateaued after $\alpha$ increased beyond a certain value. For example, in the case of PLR-R and PLR-C on the air temperature and rainfall datasets, the value of $\alpha$ had little effect when $\alpha \ge 0.25$ and $\alpha > 0.5$ respectively. In all cases, this was attributed to the cost of increasing the number of regions being too great, and the benefits of increasing the model complexity of the regions being insufficient to lower the objective function value. Furthermore, it was noted that as $\alpha$ was increased, increasing the number of clusters and regions to capture the spatio-temporal variance was not worthwhile, and so only a single region with a simple model was output. The exceptions to this were DCT and DTR modelling on the air temperature dataset and DCT modelling on the traffic dataset. As the value of $\alpha$ increased the NRMSE observed also increased and did not appear to plateau.

The range of NRMSE and storage ratios also differed between datasets for the same modelling technique and value of $\alpha$. The mean storage used was highest for the traffic dataset and lowest for the rainfall dataset. Similarly, the NRMSE achieved was highest for the traffic dataset and lowest for the rainfall dataset. We attribute this to the high temporal variance of the traffic dataset compared to the air temperature and rainfall datasets, and that the majority of rainfall instances had the same feature value (0mm rainfall). Furthermore, for the traffic dataset, discontinuities in space resulted in boundaries between regions, often yielding a higher number of regions overall. This is discussed further in Section \ref{sec:properties-discovered}. 

Finally, the relationship between storage and NRMSE differed between the modelling techniques tested. In most cases, as $\alpha$ increased, the storage used decreased in an exponential manner. However, the NRMSE appeared to follow a log-like pattern in some cases (PLR on temperature data, DTR on rainfall and traffic), quadratic pattern in some cases (DCT on traffic), arctan pattern in some cases (DCT on temperature, PLR on traffic) and a flat or linear pattern in others (DTR on temperature, PLR and DCT on rainfall). This suggests that the value of $\alpha$ used should be chosen after initial investigation of the data and, since similar relationships between NRMSE and storage ratio were observed across all samples of the same dataset given a particular modelling technique, the relationship between NRMSE, storage ratio and $\alpha$ is predictable once a subset of data samples have been tested.

\subsection{Choice of Modelling Technique}
From the results presented in Figure \ref{fig:nrmse-storage}, several conclusions about the choice of modelling technique can be drawn. First, in most cases the choice between modelling on regions or clusters has little impact on the NRMSE introduced by the resulting reduction. The only exceptions noted are for the traffic dataset when $\alpha = 0.1$, where DTR-C reported slightly lower NRMSE than DTR-R, yet PLR-R reported slightly lower NRMSE than PLR-C.

Second, the storage used by the reduced dataset differs when modelling on regions versus clusters. In the case of the air temperature dataset, modelling on clusters yielded lower storage ratios on average than modelling on regions. For example, when $\alpha = 0.1$, DTR-R used 11.7\% of the original storage volume on average whilst DTR-C used just 5.1\%. The average number of regions output in the reduced dataset was lower when modelling on clusters (as shown in Table \ref{table:regions}), though more coefficients were stored per model when modelling on clusters. This indicates the low spatial and temporal variance of the air temperature data can be more efficiently modelled with few complex models, rather than many simple models. 

In contrast, DCT-C and DTR-C yielded a higher number of regions than DCT-R and DTR-R for the traffic dataset. More complex models were required to accurately capture the high temporal variance in the traffic data when modelling on clusters. This suggests that modelling on regions may be more efficient than modelling on clusters for datasets containing higher spatial or temporal variance, though this is more noticeable for DTR modelling than DCT. For the rainfall dataset, modelling on regions and clusters achieved approximately the same NRMSE and same storage ratio. However, we noted that when modelling on regions, each region stored one coefficient for its model, whereas when modelling on clusters, each region stored a single pointer to its cluster model. The difference in storage achieved by modelling on regions or clusters was therefore negligable for all three modelling techniques.

\begin{table}[t]
     \centering
     % \small
     \caption{Average number of regions output by $k$D-STR, given differing values of $\alpha$ and the different modelling techniques evaluated.}
     \label{table:regions}
     \resizebox{\textwidth}{!}{%
     \begin{tabular}{rcccccc|cccccc}
     \cline{2-13}
     \multicolumn{1}{r}{} & \multicolumn{6}{c|}{\textbf{Air Temperature Dataset}} & \multicolumn{6}{c}{\textbf{Traffic Dataset}}  \\
     \multicolumn{1}{r}{$\alpha$} & PLR-R      & PLR-C      & DCT-R       & DCT-C      & DTR-R      & DTR-C      & PLR-R       & PLR-C       & DCT-R       & DCT-C       & DTR-R      & DTR-C      \\ \cline{2-13} 
     0.1       & 4209   & 2683   & 8403   & 2688   & 4192     & 816     & 5163    & 4060 & 3792 & 4062      & 4037      & 4235 \\
     0.25      & 1      & 1      & 1      & 1      & 1        & 1       & 2743    & 1063    & 2638 & 768      & 2255      & 1029    \\
     0.5       & 1      & 1      & 1      & 1      & 1        & 1       & 268     & 42    & 562    & 43      & 153      & 43    \\
     0.75      & 1      & 1      & 1      & 1      & 1        & 1       & 3       & 1       & 3      & 3      & 3      & 3      \\
     0.9       & 1      & 1      & 1      & 1      & 1        & 1       & 3       & 3      & 3      & 3      & 1      & 3      \\ \cline{2-13} 
     \end{tabular}
     }\vspace{0.5cm}
     \resizebox{0.55\textwidth}{!}{%
     \begin{tabular}{rcccccc}
     \cline{2-7}
     \multicolumn{1}{r}{} & \multicolumn{6}{c}{\textbf{Rainfall Dataset}}  \\
     \multicolumn{1}{r}{$\alpha$} & PLR-R      & PLR-C      & DCT-R       & DCT-C      & DTR-R      & DTR-C      \\ \cline{2-7} 
     0.1       & 3   & 3   & 3   & 3   & 3     & 3     \\
     0.25      & 3      & 3      & 3      & 3      & 3        & 3       \\
     0.5       & 3      & 1      & 3      & 1      & 1        & 3       \\
     0.75      & 1      & 1      & 1      & 1      & 1        & 1       \\
     0.9       & 1      & 1      & 1      & 1      & 1        & 1       \\ \cline{2-7} 
     \end{tabular}
     }
\end{table}

Third, the choice between polynomial linear regression, discrete cosine modelling and decision tree regression is dependent upon the user's preferences. For example, for both the air temperature and traffic datasets, DCT modelling was shown to achieve lower or similar NRMSE compared to PLR, although this comes at the cost of increased storage overhead in many cases. However, the difference in NRMSE between the two techniques is quite small, with the largest difference occurring when $\alpha=0.1$ on the traffic dataset. In this case, the average NRMSE reported was 4.7\% for DCT-C and 9.1\% for PLR-C. For the two NRMSE values reported, the storage used was 17.7\% and 14.2\% respectively. Decision tree regression, which yields easy to interpret models that can be output as a set of if-then rules, achieved similar NRMSE values on all three datasets when compared to DCT and PLR. However, this came at the cost of an increased storage overhead in the majority of cases. Therefore, when a model that is easy to interpret is required, DTR modelling is preferable, however more efficient modelling can be achieved by using PLR and DCT.

\subsection{Comparison with Other Techniques}
\begin{figure}[t]
     \centering
     \subfloat[Air Temperature]{
          \includegraphics[height=.22\textheight]{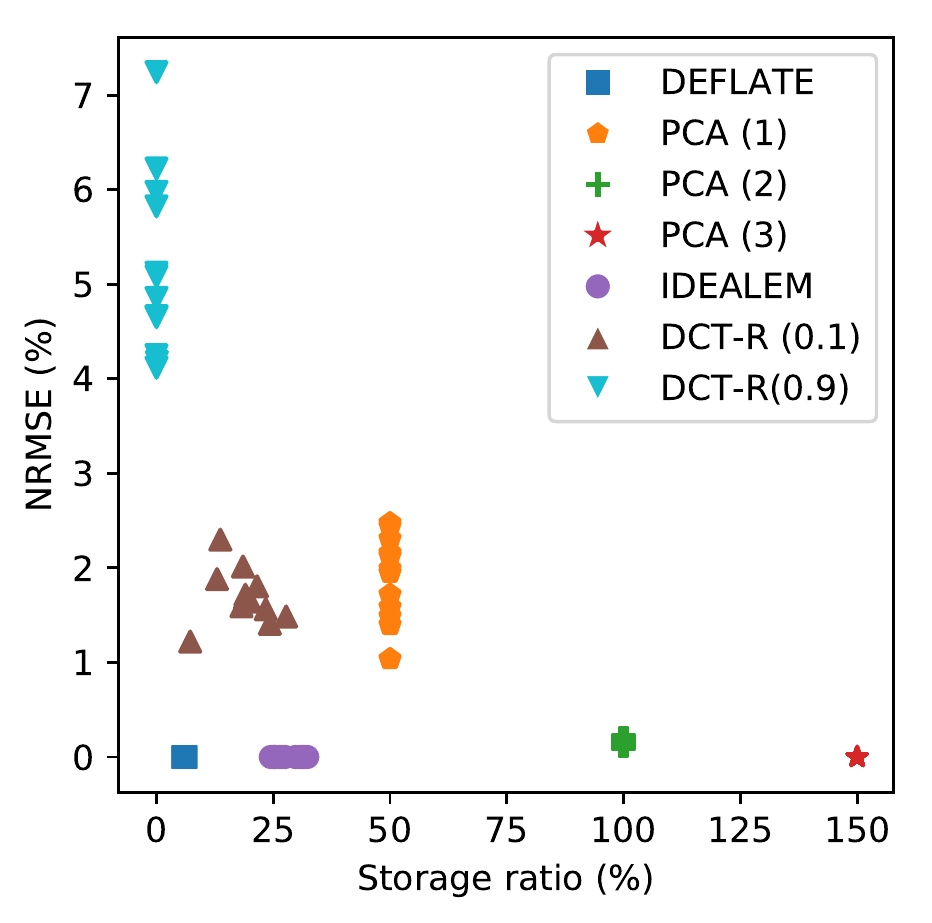}%
     }
     \subfloat[Traffic]{
          \includegraphics[height=.22\textheight]{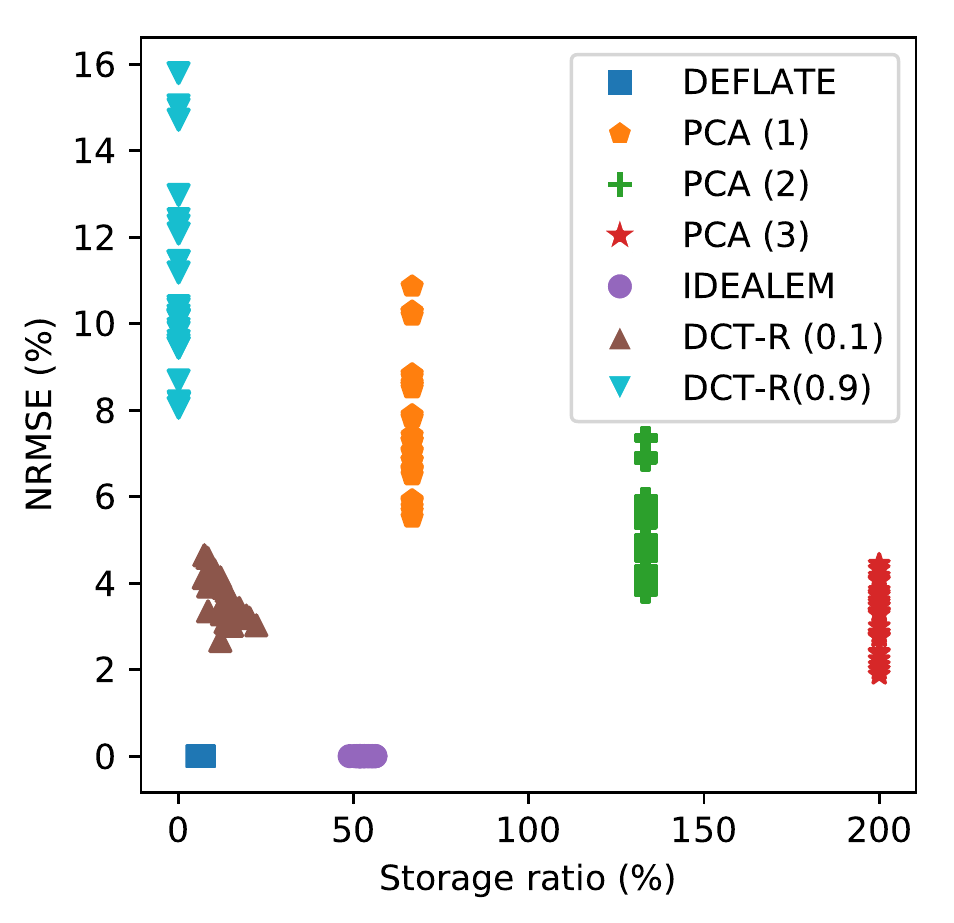}%
     } 
     \subfloat[Rainfall]{
          \includegraphics[height=.22\textheight]{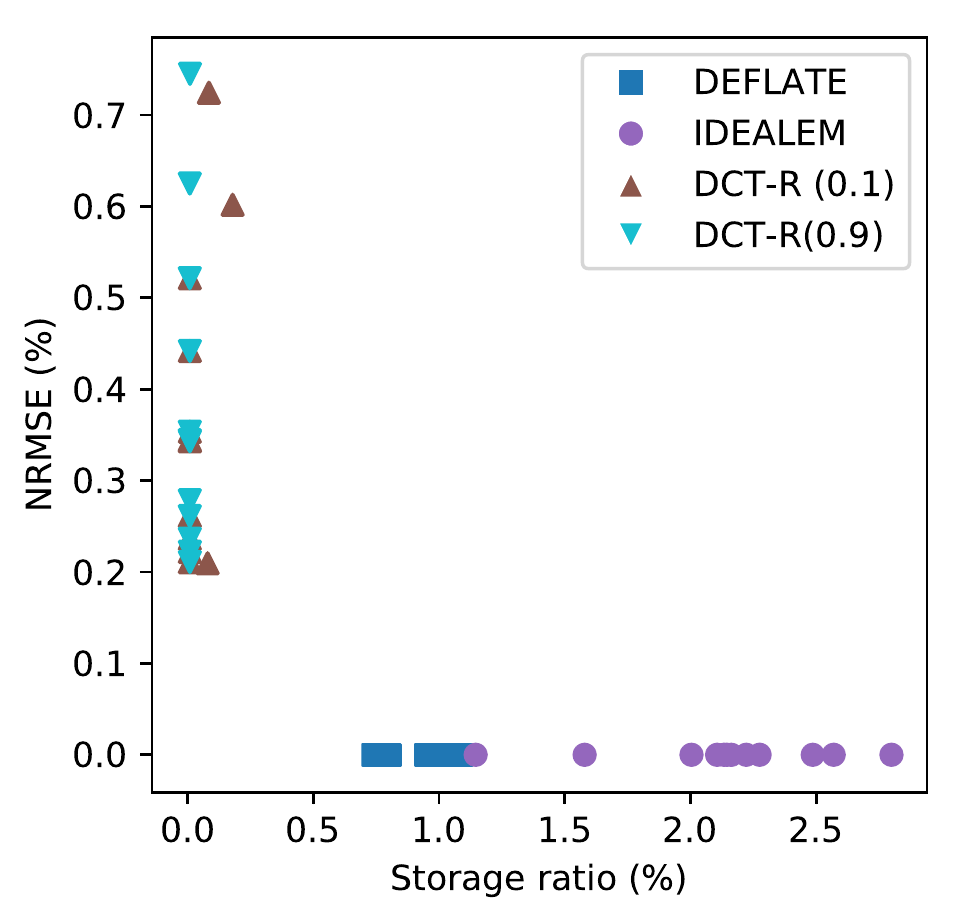}%
     }
     
     \caption{Comparison of $k$D-STR with PCA, IDEALEM and DEFLATE on the air temperature, traffic and rainfall datasets. Note PCA is not shown for the rainfall dataset, though it achieved negligible NRMSE using 25\% of the original dataset volume. For $k$D-STR, discrete cosine modelling on regions was chosen to represent the range of reduction results seen.}
   \label{fig:comparisons}
\end{figure}

As discussed in Section \ref{sec:evaluation}, we compared $k$D-STR with the IDEALEM, PCA and DEFLATE methods. 
We used DEFLATE, a lossless compression algorithm, as a benchmark to show the reduction in storage achievable whilst losing no information. 
Whilst DEFLATE is able to compress the data to a form smaller than the raw dataset, the entire dataset must be decompressed before the data can be used for analysis. 
Thus, the user must have sufficient memory to store the decompressed raw dataset and cannot use the compressed data directly. 
In comparison, $k$D-STR allows the source data to be reconstructed by inputting the desired locations and times into the model(s) of the relevant regions in the reduced dataset.
Furthermore, some analysis can be performed on the models output by $k$D-STR without needing reconstruct the data. 
Therefore, we include DEFLATE as an indicator of the potential reduction in storage that is achievable by lossless compression algorithms, rather than as a directly comparable reduction method. 
When evaluated on the datasets presented in this paper, DEFLATE reduced the data to between 0.7\% and 7.2\% of the original volume. 
DEFLATE reduced the air temperature datasets to an average of 6.0\% of the original data volume, the traffic datasets to 6.3\% and the rainfall datasets to 1.0\%, as shown in Figure \ref{fig:comparisons}.

PCA, adapted for spatio-temporal sensor data\footnote{We used the atmospheric science PCA adaptation, as discussed in \cite{demvsar2013principal}, as this reflected the nature of data collected from sensors that are spatially and temporally autocorrelated.}, was found to reduce the air temperature datasets to 50\% of the original dataset volume when 1 principal component was selected, and 100\% (meaning no reduction in storage) when 2 principal components were selected. An average of 1.9\% and 3.95$\times 10^{-17}$\% NRMSE was achieved respectively. On the traffic dataset, 1 principal component was found to produce an average error of 7.6\% whilst using 67\% of the original data volume, 2 principal components were found to produce an average error of 5.2\% error whilst using 133\% of the original data volume and 3 principal components again produced an average error of 3.2\% error whilst using 200\% of the original data volume. Whilst not shown on Figure \ref{fig:comparisons}(c), 1 principal component was also found to produce negligible NRMSE on the rainfall dataset whilst using 25\% of the original data volume. 

IDEALEM was found to produce negligible NRMSE across all of the datasets evaluated. IDEALEM reduced the air temperature datasets to between 24.5\% and 32.3\% of the original data volume, the traffic datasets to between 48.8\% and 56.2\% and the rainfall datasets to between 1.1\% and 2.8\% of the original data volume. 

In comparison to these results, $k$D-STR was found to reduce the datasets to smaller volumes than IDEALEM and PCA, achieving storage ratios more similar to DEFLATE for both the air temperature and traffic datasets. 
Results for $k$D-STR using DCT modelling on regions (DCT-R) are shown to indicate the performance achieved by $k$D-STR compared to other techniques. 
When $\alpha = 0.1$, indicating a preference for minimal introduced error at the cost of increased storage used, $k$D-STR reduced the air temperature datasets to between 7.2\% and 27.8\%, the traffic datasets to between 7.2\% and 22.2\%, and the rainfall datasets to between 0.08\% and 0.18\% of the original dataset volumes. 
This was achieved by introducing an NRMSE of between 1.2\% and 2.3\%, 2.7\% and 4.6\%, and 0.2\% and 0.8\% accordingly. 
When $\alpha = 0.9$, indicating a preference for minimal storage at the cost of increased NRMSE, all datasets were reduced to less than 0.16\% of the original volume. 
These reductions resulted in a NRMSE of between 4.1\% and 7.2\% for the air temperature datasets, 8.1\% and 15.8\% for the traffic datasets, and 0.2\% and 0.8\% for the rainfall datasets.

These results indicate that $k$D-STR is able to reduce the datasets evaluated to smaller storage volumes than IDEALEM and PCA, often with quite significant improvements. However, $k$D-STR may incur increased information loss compared to the other techniques evaluated.

\subsection{Choice of Spatial Referencing System}
Some datasets can be referenced using multiple spatial domains. For example, transportation datasets that are sourced from roads can be referenced using a 2-dimensional spatial domain or a 1-dimensional spatial domain (or linear referencing system). When such datasets can be reduced using either of their spatial referencing systems (SRSs), the user must decide which they wish to use. Therefore, we compared the reduction of the 28 traffic dataset samples using $k$D-STR when 1-dimensional and 2-dimensional spatial referencing systems can be used, and refer to these as $k=2$ and $k=3$ respectively.

\begin{sidewaysfigure}
     \centering
     \vspace{7cm}
     \subfloat[NRMSE]{
          \includegraphics[width=0.87\textheight]{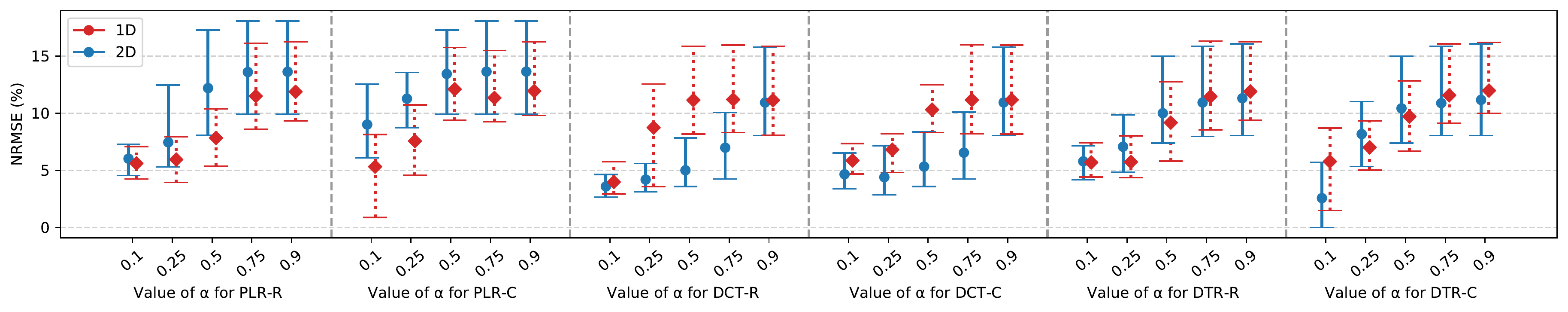}%
     }

     \subfloat[Storage used]{
          \includegraphics[width=0.87\textheight]{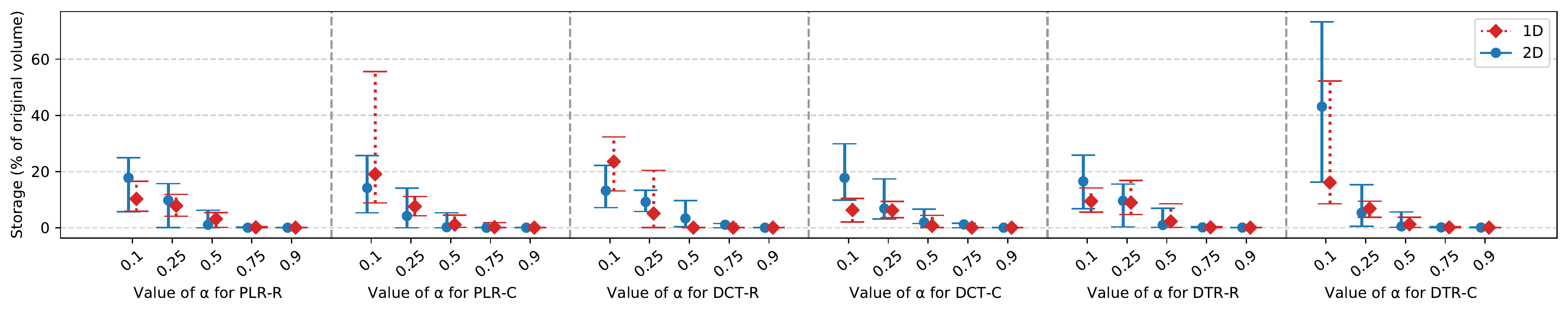}%
     }

     \subfloat[Number of regions output]{
          \includegraphics[width=0.87\textheight]{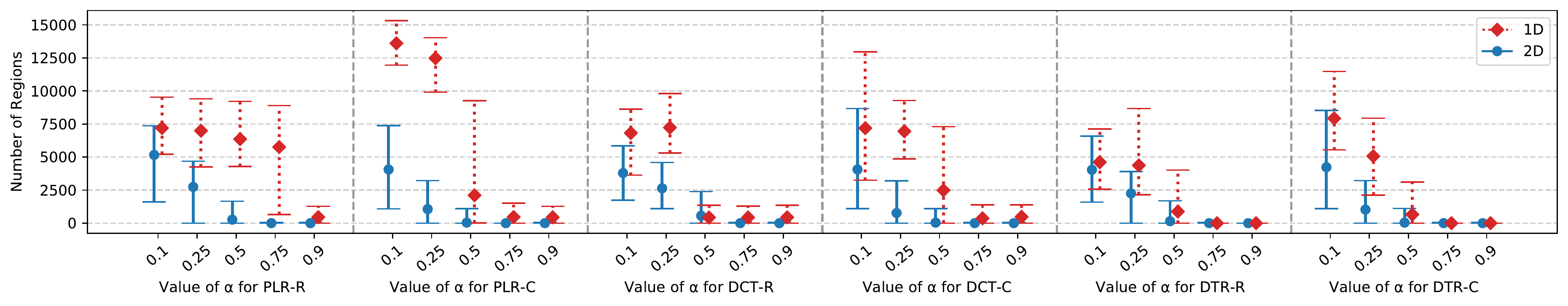}%
     }
     
     \caption{Error introduced, storage used and number of regions output by $k$D-STR on the traffic dataset using a 1-dimensional and 2-dimensional spatial referencing system (SRS). Each sub-figure shows the results of $k$D-STR using the six modelling techniques evaluated for varying values of $\alpha$.}
   \label{fig:traffic-comparison}
   \vspace{-7cm}
\end{sidewaysfigure}

A comparison of the NRMSE incurred, storage used and number of regions output by \textit{k}D-STR when applied to the traffic dataset using 1-dimensional and 2-dimensional spatial domains can be seen in Figure \ref{fig:traffic-comparison}. The relationship between NRMSE and storage used when $k=2$ and $k=3$ was similar for all modelling techniques used. However, the number of regions output when $k=2$ was consistently higher than when $k=3$. In some cases, for example DTR-R modelling, the number of regions output differed slightly whilst in others, for example PLR-C modelling, the number of regions output differed significantly. Furthermore, in some cases, notably PLR-R and DTR-R modelling, the higher number of regions used when $k=2$, combined with fewer model coefficients stored per region, led to a lower or similar average NRMSE with a lower quantity of storage used. In most other cases a lower quantity of storage used correlated with a higher NRMSE achieved and a higher quantity of storage used correlated with a lower NRMSE achieved.

\subsection{Spatial and Temporal Properties Discovered by $k$D-STR}
\label{sec:properties-discovered}
The partitioning method used by $k$D-STR was found to identify several patterns and nuances within each of the three datasets tested. 
For example, for the air temperature dataset, several regions were created per day but few regions were created in space. 
We attributed this to the higher temporal variance in the dataset compared to its spatial variance.
A similar partitioning in time occurred for the traffic dataset, however main carriageway and slip-road sensors were consistently placed in different regions during the daytime.
This distinction was a result of the spatial discontinuity between the main carriageway and slip-roads which resulted in significant differences in the feature values recorded by the two types of sensor. 
From these two datasets, we therefore concluded that more regions were created in the spatial domain when the spatial domain has a higher variance than the temporal domain, and vice versa.

However, this conclusion did not hold for the rainfall dataset. 
Whilst the rainfall dataset experienced discontinuities at times of rainfall versus no rainfall, a maximum of 3 regions were introduced for each of the $\alpha$ values tested.
We attributed this to the cost of increasing the number of regions being too high to be chosen by $k$D-STR. 
We therefore suggest that when the data contains a high enough number of discontinuities, for example a rainfall dataset that contains many occurrences of rainfall that are limited in their spatial area and duration in time, $k$D-STR will create few regions and attempt to model these discontinuities with complex models.

\subsection{Discussion}
From the results presented in this section, several observations can be drawn. First, $k$D-STR is shown to be effective in reducing a spatio-temporal dataset to a set of regions of similar instances in space and time and forming a model of the instances within each of these regions. Furthermore, the parameter $\alpha$ is shown to be effective at allowing the user to decide between minimising the information lost and minimising the storage used. This is shown for a variety of datasets that exhibit different spatio-temporal characteristics, including different rates of variance in space and time, and different rates of continuous variability and sudden changes (events).

Second, $k$D-STR is shown to perform similarly when 1-dimensional and 2-dimensional spatial referencing systems are used for the same dataset. A similar relationship between NRMSE and storage is achieved both when $k=2$ and $k=3$, again showing that $\alpha$ can be used to control the rate of information lost and storage used. We believe this to be further indication of the generality of spatio-temporal datasets that can be reduced by $k$D-STR.

Third, the results shown in Section \ref{sec:results-alpha} indicate that the rate of information lost and storage used are more correlated with the dataset being reduced than the method used to model the instances within each region. This allows the user to select the modelling technique that is most applicable for their general form of analysis from a suite of possible modelling techniques. For example, if the user wishes to understand the relationships between feature values and the spatio-temporal dimensions, PLR may be preferable. However, when minimising NRMSE is more important it may be more beneficial to use DCT. Furthermore, when a more interpretable model is required for each region, DTR can be used.

Finally, $k$D-STR is shown to achieve similar storage ratios to DEFLATE and to achieve smaller storage ratios than both PCA and IDEALEM. Whilst the lossless compression algorithm DEFLATE achieves 0\% NRMSE, it does not permit analysis of the dataset without first requiring decompression back into the original dataset. In comparison, $k$D-STR permits multiple types of analysis on the reduced datasets without requiring a transformation back from the reduced state. $k$D-STR is also able to take advantage of the spatial nature of spatio-temporal data, unlike IDEALEM, and does not require a mapping from the stored models back to the original feature space like PCA.

\section{Conclusion}
\label{sec:conclusion}

In this paper, we introduced $k$D-STR, a method for reducing spatio-temporal datasets in a manner that permits multiple types of analysis on the resulting reduced dataset. $k$D-STR overcomes the issue of region forming present in 2D-STR when the number of spatial dimensions is greater than 1. $k$D-STR uses hierarchical partitioning to decompose space and time into regions of similar instances. Then, each region is modelled using an appropriate technique, thus reducing the original dataset to a set of regions and their models. This process reduces the quantity of data that needs to be processed for analysis and answering questions.

We have demonstrated the effectiveness of $k$D-STR in achieving an average reduction of 99.7\% in storage on the datasets used for evaluation when a reduction in dataset volume is preferred, whilst achieving an average NRMSE of 7.6\%. Further, we have demonstrated $k$D-STR achieves a reduction of 86.5\% in storage for the same datasets, whilst achieving an average NRMSE of 3.5\% when a minimal introduced error is preferred. The number of regions output is significantly smaller than the number of instances in the original datasets, reducing the quantity of data to be processed for analysis.

In comparison to other techniques, $k$D-STR is found to reduce a dataset to sizes similar to DEFLATE whilst achieving NRMSE error rates similar to IDEALEM and PCA. Therefore, $k$D-STR is shown to perform comparably to state of the art techniques whilst permitting multiple types of analysis to be performed on the reduced dataset. Furthermore, the level of reduction can be controlled using the parameter $\alpha$, allowing the user to choose the  level of reduction they wish to perform.

We believe the investigation of dataset reduction methods to be an important avenue of research. Methods like $k$D-STR, which permit multiple types of analysis to be performed on the reduced dataset, are important in allowing data scientists to work in a more efficient manner as the quantity of data in spatio-temporal datasets increases. $k$D-STR has been shown to be effective at reducing data with different spatial and temporal variances and trends (such as air temperature and traffic data), as well as data that exhibits frequent discontinuities in space and time (such as rainfall data). However, we may take further advantace of the characteristics of such datasets. For example, datasets which exhibit oscillations will be broken into multiple regions by the partitioning stage of $k$D-STR. Yet, by partitioning the spatio-temporal space into regions of instances that exhibit clear trends or patterns, we may be able to further reduce such types of data. Future extensions of this work will also investigate the reduction of data that exhibits autocorrelations in higher numbers of dimensions, such as high-dimensional simulation data, and investigate the process of reducing multiple linked datasets at the same time. Furthermore, future extensions may examine the effect of reduction on linking spatio-temporal datasets and prioritise the performance of predictive models built using reduced datasets, rather than prioritising the retention of nuances present in the raw data. Finally, in this work we focused on modelling data for which a maximum of one instance was recorded at any given location and time. Future work may look at reducing datasets where multiple instances may be recorded at the same time and location, perhaps using modelling techniques that are built for multiple dependent or response values for the same independent or predictor values.

%
% The acknowledgments section is defined using the "acks" environment (and NOT an unnumbered section). This ensures
% the proper identification of the section in the article metadata, and the consistent spelling of the heading.
\begin{acks}
The lead author gratefully acknowledges funding by the UK Engineering and Physical Sciences Research Council (grant no. EP/L016400/1), the EPSRC Centre for Doctoral Training in Urban Science, and funding from TRL.
\end{acks}

\bibliographystyle{ACM-Reference-Format}
\bibliography{Bib20181105}

\end{document}